\documentclass[
  aps,
  prd,
  twocolumn,
  superscriptaddress,
  nofootinbib,
  longbibliography,
]{revtex4-2}

\usepackage[utf8]{inputenc}
\usepackage[T1]{fontenc}
\usepackage{graphicx}
\usepackage{dcolumn}
\usepackage{bm}
\usepackage{amsmath}
\usepackage{amssymb}
\usepackage{amsthm}
\usepackage{mathrsfs}
\usepackage{tikz}
\usetikzlibrary{arrows.meta}
\usepackage[table]{xcolor}
\usepackage{etoolbox}
\usepackage[colorlinks=true,allcolors=blue]{hyperref}
\usepackage{orcidlink}

\graphicspath{{figs/}}

\AtBeginDocument{%
  \setcounter{topnumber}{3}%
  \setcounter{bottomnumber}{2}%
  \setcounter{totalnumber}{4}%
  \setcounter{dbltopnumber}{3}%
}

\theoremstyle{plain}

\theoremstyle{definition}

\theoremstyle{remark}

\begin{document}

\title{Unified matched-filter and excess-power pipeline for gravitational-wave detector characterization}

\author{Zach Yarbrough \orcidlink{0000-0002-9825-1136}}
\affiliation{Department of Physics and Astronomy, Louisiana State University, Baton Rouge, LA 70803, USA}

\author{Olivia Godwin \orcidlink{0000-0002-7489-4751}}
\affiliation{LIGO Laboratory, California Institute of Technology, MS 100-36, Pasadena, California 91125, USA}
\affiliation{Department of Physics, The Pennsylvania State University, University Park, PA 16802, USA}
\affiliation{Institute for Gravitation and the Cosmos, The Pennsylvania State University, University Park, PA 16802, USA}

\author{Derek Davis \orcidlink{0000-0001-5620-6751}}
\affiliation{University of Rhode Island, Kingston, RI 02881, USA}

\author{Gabriela Gonz\'alez \orcidlink{0000-0003-0199-3158}}
\affiliation{Department of Physics and Astronomy, Louisiana State University, Baton Rouge, LA 70803, USA}

\date{\today}

\begin{abstract}
We present \textsc{sgnax}, an open-source pipeline for gravitational-wave detector characterization that delivers both matched-filter and excess-power transient triggers from a single streaming dataflow graph.
Built on the Stream Graph Navigator (\textsc{sgn}) framework, \textsc{sgnax} unifies the multi-rate sine-Gaussian matched-filter search of \textsc{snax} and the multi-resolution Q-transform excess-power search of \textsc{omicron} in one \textsc{python}-native package with a shared configuration and output schema.
Strain and auxiliary channels from a single interferometer can be analyzed as parallel branches of one graph; because the branches share one data-read and whitening front end, the per-channel processing cost falls as channels are added---by a factor of $2.8$ between one and $32$ channels for the Q-transform---and a full day of $16\,\mathrm{kHz}$ strain is analyzed in $14$ minutes of wall-clock time.
In a $512\,\mathrm{s}$ archival-data benchmark, four processes sharing one NVIDIA V100 analyze $792$ production auxiliary channels with the Q-transform at $1.11\times$ real-time streaming throughput.
Selectable data sources span offline frame caches, shared-memory buffers, and the \textsc{arrakis} distribution service, with one configuration serving offline and online operation.
The streaming matched filter carries no per-chunk buffering and delivers triggers at a measured end-to-end latency of about five seconds, while the Q-transform reaches low-latency operation within tens of seconds of acquisition, well ahead of the cadence at which production \textsc{omicron} deployments deliver triggers in practice.
Injection campaigns validate both modes, recovering $99.4\%$ of recoverable sine-Gaussian injections with parameters within the expected template mismatch, and broadband white-noise-burst recovery consistent with the established event-trigger generators.
On 24 hours of archival LIGO strain, \textsc{sgnax} reproduces the production \textsc{omicron} trigger population to $0.1\%$ in total trigger count, with per-trigger SNRs matched at the percent level in every frequency band and the residual differences localized to a documented difference in whitening adaptation during noise transients; on production auxiliary channels, its dense feature streams recover the \textsc{snax} loud-feature population with $80\%$ per-bin coincidence, a comparison limited less by the reimplementation than by the production reference itself.
Indeed, the injection-calibrated reimplementation uncovers a multiband amplitude error in the production \textsc{snax} deployment that inflates its reported SNRs below $25.6\,\mathrm{Hz}$ by factors of $\sqrt{2}$--$2$, for which we give the mechanism and the correction.
\end{abstract}

\maketitle

\section{Introduction}\label{sec:introduction}

The LIGO--Virgo--KAGRA (LVK) network of ground-based interferometers~\cite{aasi2015aligo, acernese2015advirgo, kagra2019} has detected hundreds of astrophysical gravitational-wave signals~\cite{abbott2016, abbott2021gwtc3, abac2025gwtc4, abac2026gwtc5}.
Detecting these transients requires distinguishing genuine astrophysical signals from a population of non-stationary instrumental and environmental artifacts---``glitches''---that contaminate the strain time series~\cite{abbott2016detchar, davis2021detchar, soni2025o4adetchar, glanzer2026o4bcdetchar}.
Detector characterization workflows rely on continuous transient extraction not only from the strain channel itself, but from the tens of thousands of \emph{auxiliary channels}---time series recorded by the sensors that monitor the instrument and its environment~\cite{mciver2019diagnostic, robinet2020omicron}.
Many of the transient-noise classes of greatest concern for detector characterization---scattered-light glitches foremost among them~\cite{soni2021scattering}---populate the low-frequency band below roughly $100\,\mathrm{Hz}$, which is consequently a primary focus of characterization effort~\cite{davis2021detchar}.
Two software pipelines have performed the bulk of this work in recent years.
\textsc{omicron}~\cite{robinet2020omicron, robinet2018technote, omicron_repo}, a C++ implementation of the constant-Q transform~\cite{brown1991qtransform} originally developed in 2012, is the primary trigger generator used across the international gravitational-wave detector network for detector characterization and event validation.
In the LIGO production configuration considered here, \textsc{omicron} analyzes strain together with approximately $800$ auxiliary channels per detector.
\textsc{snax}~\cite{snax_repo, godwin2020thesis}---``Stream-based Noise Acquisition and eXtraction''---is a \textsc{python} package, separated from the \textsc{gstlal} framework~\cite{messick2017, cannon2021gstlal} into a standalone package in 2022, that performs multi-rate sine-Gaussian matched filtering on auxiliary channels~\cite{godwin2020thesis}.
The production use of \textsc{snax} has been confined to auxiliary channels and its matched-filter analysis; it did not provide a Q-transform mode.
The two tools target overlapping problems with distinct algorithmic approaches and have evolved independently within the gravitational-wave software stack.

In this paper we describe \textsc{sgnax}, a single \textsc{python} package that subsumes both functionalities by reimplementing them as elements of the Stream Graph Navigator (\textsc{sgn}) dataflow framework~\cite{huang2026sgn}.
\textsc{sgnax} provides two analysis modes through its command-line entry points.
The first, \texttt{sgnax-extract}, performs the multi-rate sine-Gaussian matched-filter search inherited from \textsc{snax}, evaluating correlations on either CPU or GPU through the \textsc{pytorch} backend of the \textsc{sgn-ts} time-series library~\cite{huang2026sgn, paszke2019pytorch}; a companion entry point, \texttt{sgnax-timeseries}, runs the same search threshold-free and emits \textsc{snax}'s dense regularly-sampled feature-timeseries data product (Sec.~\ref{sec:algorithms:matched_filter}).
The second, \texttt{sgnax-qtransform}, performs the bisquare-windowed Q-transform excess-power search of \textsc{omicron}, including the same time--frequency tiling, whitening normalization, and signal-to-noise estimator as in Robinet \emph{et al.}~\cite{robinet2020omicron}.
Through a unified configuration object (\texttt{DataSourceInfo}), both entry points consume any of three data-source backends: offline frame caches, the low-latency shared-memory buffer\footnote{Known internally as \texttt{/dev/shm}, after the path at which the shared-memory partition is mounted at the detector sites.}, and the \textsc{arrakis} distribution service~\cite{arrakis}.
Both modes can analyze either strain or auxiliary channels.
Our planned deployment uses the Q-transform on strain and the auxiliary-channel list to provide \textsc{omicron}-equivalent trigger products, while the matched-filter mode supplies auxiliary-channel triggers and features on the shorter timescales required by tools such as \textsc{idq}~\cite{essick2021idq, Huxford:2024vdt}.
The pipeline stores triggers in a single \textsc{hdf5} schema common to both algorithms.

The motivation for the rewrite is fourfold.
First, the operational cost of maintaining \textsc{snax} as a thin \textsc{python} layer over \textsc{gstlal}'s \textsc{gstreamer}-1.0 dataflow has grown with each upstream release, and the \textsc{sgn} framework~\cite{huang2026sgn}---developed originally for the inspiral-search pipeline \textsc{sgnl}~\cite{cannon2021gstlal}---offers a typed, native-\textsc{python} alternative whose elements can be composed without a C-level pipeline backbone.
Second, \textsc{omicron} requires a heavy set of external libraries and two separate build systems---detailed in Appendix~\ref{sec:appendix:implementation:predecessors}---which pose a persistent challenge for non-expert users.
This motivates an in-process \textsc{python} implementation that can be deployed alongside other detector-characterization tools without leaving the \textsc{python} interpreter.
Third, neither predecessor processes multiple auxiliary channels through a single coherent dataflow graph: \textsc{snax} fans out via \textsc{htcondor} jobs over channel subsets~\cite{snax_repo}, and \textsc{omicron}---itself commonly distributed over \textsc{htcondor} by the \textsc{pyomicron} wrapper~\cite{pyomicron_repo}---processes channels sequentially within each analyzed data segment~\cite{robinet2018technote}.
\textsc{sgnax} processes an arbitrary list of channels from a single interferometer through one pipeline, with the channels analyzed concurrently as parallel branches of a single dataflow graph rather than as separate jobs.
Fourth, although \textsc{omicron} supports an online mode with minutes-scale trigger delivery~\cite{robinet2020omicron}, production deployments in recent observing runs have in practice delivered Q-transform triggers to detector commissioning consoles on substantially longer cadences set by the batch workflow; a \textsc{python}-native pipeline that consumes the low-latency shared-memory and \textsc{arrakis} data sources through the same configuration object used for offline runs makes Q-transform triggers available within tens of seconds, without a separate online infrastructure.

Beyond this consolidation, hosting the two analyses on a common data path and configuration makes possible a set of methodological comparisons that the historically separate, independently-configured tools did not readily admit, and which form the quantitative core of this paper.
We compare matched-filter and excess-power detection like-for-like on a shared injection set (Sec.~\ref{sec:results:modes}).
We extend the classic event-trigger-generator validation of McIver and Godwin~\cite{mciver2015thesis, godwin2020thesis} to broadband white-noise bursts, showing that a configurable cluster-SNR estimator recovers more of their spread power than the single-tile statistic used by the production trigger generators (Sec.~\ref{sec:results:wnb}).
We characterize the detection efficiency of both modes against their false-alarm rate (Sec.~\ref{sec:results:roc}), and we measure the online trigger latency of each mode directly (Sec.~\ref{sec:results:latency}).
We also measure the capacity of a single data-center GPU for the multi-channel Q-transform workload, measuring throughput above real time for $792$ production auxiliary channels on one device over a $512\,\mathrm{s}$ benchmark (Sec.~\ref{sec:results:gpu}).
Finally, we compare both modes head-to-head with their predecessors on production data---against \textsc{omicron} on archival strain (Sec.~\ref{sec:results:omicron}) and against the \textsc{snax} production features on auxiliary channels (Sec.~\ref{sec:results:snax})---where the injection-calibrated reimplementation surfaces a low-frequency SNR-scale excess in the production \textsc{snax} deployment.

The remainder of this paper is organized as follows.
Section~\ref{sec:predecessors} reviews \textsc{snax} and \textsc{omicron} in enough detail to allow a precise comparison with \textsc{sgnax}.
Section~\ref{sec:architecture} describes the \textsc{sgn}-based dataflow architecture, the three data-source backends, and the trigger output schema.
Section~\ref{sec:algorithms} describes the two analysis algorithms, including the bank-spacing metric for the matched-filter search and the bisquare-window normalization for the Q-transform, and identifies the points at which \textsc{sgnax} differs from its predecessors.
Section~\ref{sec:results} validates both analyses against injections, compares them with the predecessor pipelines on archival strain and production auxiliary channels, and benchmarks their runtime and online latency.
Section~\ref{sec:discussion} discusses the migration path from existing \textsc{snax} and \textsc{omicron} workflows and the operational role we anticipate for \textsc{sgnax}.

\section{Predecessor pipelines}\label{sec:predecessors}
\subsection{\textsc{snax}: stream-based sine-Gaussian extraction}\label{sec:predecessors:snax}

\textsc{snax} performs a multi-rate sine-Gaussian matched-filter search on auxiliary channels; it was originally part of the \textsc{gstlal-burst} package and became a standalone package in 2022~\cite{snax_repo}.
Its production analyses were restricted to auxiliary channels and used matched filtering; \textsc{snax} did not include a Q-transform analysis.
For a time series $a(t)$ whitened to unit-variance, \textsc{snax} computes the correlation
\begin{equation}\label{eq:matched_filter}
  \rho_\theta(t) = \int a(t')\,h_\theta(t' - t)\,\mathrm{d}t'
\end{equation}
against a bank of unit-norm sine-Gaussian templates parameterized by $(f, Q)$,
\begin{equation}\label{eq:sine_gaussian}
  h_{f,Q,\varphi}(t) = \mathcal{N}_{f,Q}\,
    \cos\!\left(2\pi f t + \varphi\right)\,
    \exp\!\left[-\frac{(2\pi f t)^2}{2 Q^2}\right],
\end{equation}
with both quadrature phases $\varphi \in \{0, \pi/2\}$ and a normalization $\mathcal{N}_{f,Q}$ chosen so that $\|h_{f,Q,\varphi}\|_2 = 1$.
The bank is populated by a mismatch-driven log-spaced grid in $(f, Q)$ derived from the sine-Gaussian metric~\cite{chatterji2005}; the spacing is recapitulated in Appendix~\ref{sec:appendix:bank_spacing}.
Triggers are extracted by combining the quadrature outputs into the SNR estimator
\begin{equation}\label{eq:snr_iq}
  \rho_{f,Q}(t) = \sqrt{\rho_{f,Q,0}^2(t) + \rho_{f,Q,\pi/2}^2(t)},
\end{equation}
applying a per-window peak finder, and writing to disk the resulting time, SNR, frequency, $Q$, phase, and duration.

The \textsc{snax} pipeline is implemented as a \textsc{python} layer over the \textsc{gstreamer}-1.0 dataflow of \textsc{gstlal}~\cite{snax_repo, cannon2021gstlal}.
The template bank is partitioned across power-of-two sampling rates, each handled by a dedicated set of pipeline elements, and multi-channel coverage is achieved at the workflow level by distributing the channel list among parallel \textsc{htcondor} jobs whose output files are concatenated downstream.
The pipeline is mature and stable, with eleven command-line tools spanning offline batch analysis, low-latency online operation via \textsc{kafka}, and real-time monitoring~\cite{snax_repo}.
Its dependence on the underlying \textsc{gstreamer} framework, however, carries the operational cost that motivates the present implementation; the element-level structure of the pipeline and the maintenance burden it imposes are detailed in Appendix~\ref{sec:appendix:implementation:predecessors}.

\subsection{\textsc{omicron}: the Q-transform pipeline}\label{sec:predecessors:omicron}

\textsc{omicron}~\cite{robinet2020omicron, robinet2018technote, omicron_repo} is a C++ implementation of the constant-Q transform~\cite{brown1991qtransform} originally developed by Robinet \emph{et al.} in 2012 and continuously maintained since.
For a whitened time series $x^{\rm wh}(t)$, the Q-transform is~\cite{robinet2020omicron}
\begin{equation}\label{eq:qtransform}
  X(\tau, f, Q) = \int_{-\infty}^{+\infty}
    x^{\rm wh}(t)\,
    w(t - \tau, f, Q)\,
    e^{-2i\pi f t}\,\mathrm{d}t,
\end{equation}
with $w$ an analysis window of duration $\sigma_t = Q/(\sqrt{8\pi}\,f)$ that varies inversely with the central frequency $f$.
The parameter space $(\tau, f, Q)$ is partitioned into logarithmically-spaced Q-planes; within each plane, frequency rows are log-spaced and time bins are linearly spaced~\cite{chatterji2004multiresolution, chatterji2005, robinet2020omicron}.
After normalizing such that the expectation value of $|X|^2$ for stationary Gaussian noise equals~2, an excess-power signal-to-noise estimator is defined as~\cite[Eq.~2]{robinet2020omicron}
\begin{equation}\label{eq:omicron_snr}
  \hat\rho(\tau, f, Q) = \begin{cases}
    \sqrt{|X^{\rm wh}(\tau, f, Q)|^2 - 2}
      & \text{if } |X^{\rm wh}|^2 \geq 2, \\
    0 & \text{otherwise.}
  \end{cases}
\end{equation}
Tiles whose SNR exceeds a configured threshold are clustered in time and saved as triggers.

\textsc{omicron} occupies a central role in gravitational-wave detector characterization workflows~\cite{robinet2020omicron, mciver2019diagnostic}, analyzing both strain and, in the LIGO production configuration considered here, approximately $800$ auxiliary channels per detector.
Its triggers feed downstream tools including \textsc{gravity~spy}~\cite{zevin2017gravityspy, glanzer2023gravityspy}, \textsc{hveto}~\cite{smith2011hveto}, machine-learning glitch classifiers~\cite{biswas2013mla}, and event-validation infrastructure such as the data-quality report~\cite{davis2026dqr}.
The pipeline is invoked through a parameter-file-driven command-line interface; output triggers are serialized as \textsc{root} \texttt{TTree} objects~\cite{root1997}.

The implementation is, however, difficult to deploy outside the computing environments in which it has traditionally been supported, owing to a heavy external dependency tree~\cite{root1997, gwollum_repo, frigo2005fftw} and a dual \textsc{cmt}/\textsc{cmake} build system; these details, together with the \textsc{pyomicron} wrapper~\cite{pyomicron_repo} used to drive it at the LIGO sites, are collected in Appendix~\ref{sec:appendix:implementation:predecessors}.

\subsection{Common motivation for a unified rewrite}\label{sec:predecessors:motivation}

\textsc{snax} and \textsc{omicron} extract transients for detector characterization through different approaches: a parametric matched filter against an explicit sine-Gaussian template bank, and a non-parametric excess-power tiling derived from the Q-transform.
Their production channel coverage also differs: \textsc{snax} analyzes auxiliary channels, while \textsc{omicron} analyzes both strain and auxiliary channels.
For the auxiliary-channel use case, the two approaches are largely complementary, and detector-characterization tools have come to rely on triggers from both.
Our goal in developing \textsc{sgnax} was to make this complementarity available within a single \textsc{python}-native pipeline that shares one dataflow graph, configuration, output schema, and dependency footprint across the two algorithms, and that expresses multi-channel processing within the dataflow graph itself rather than relying solely on a job-management system to distribute the channel list.

\section{Architecture}\label{sec:architecture}
\subsection{The \textsc{sgn} dataflow framework}\label{sec:architecture:sgn}

\textsc{sgnax} is built on the Stream Graph Navigator (\textsc{sgn}) dataflow framework~\cite{huang2026sgn}, the same framework that underlies the inspiral-search pipeline \textsc{sgnl} and a growing collection of gravitational-wave low-latency tools.
An \textsc{sgn} pipeline is a directed graph of \emph{elements} connected through named \emph{pads}.
Three element categories are distinguished: \emph{sources}, which produce frames at the head of the graph; \emph{transforms}, which consume one or more input frames and produce output frames; and \emph{sinks}, which terminate a branch by writing to disk, network, or memory.
Frames are typed: \texttt{TSFrame} carries time-series samples (provided by \textsc{sgn-ts}~\cite{huang2026sgn}), and \texttt{EventFrame} carries trigger records.
End-of-stream propagation is handled by the framework rather than by ad-hoc signaling, in contrast to the \textsc{gstreamer}~\cite{cannon2021gstlal} backbone of \textsc{snax}.

\textsc{sgnax} draws elements from four \textsc{sgn} ecosystem packages.
\textsc{sgn}~\cite{huang2026sgn} provides the \texttt{Pipeline}, \texttt{SourceElement}, \texttt{TransformElement}, and \texttt{SinkElement} classes.
\textsc{sgn-ts}~\cite{huang2026sgn} provides time-series elements including \texttt{Resampler}, \texttt{Whiten}, and \texttt{Correlate}; the last of these dispatches to a \textsc{pytorch}~\cite{paszke2019pytorch} backend that runs transparently on CPU or GPU.
\textsc{sgn-ligo}~\cite{sgnligo_repo} provides LIGO-specific frame-cache and shared-memory sources, including the \texttt{/dev/shm} multicast reader used in low-latency operation.
\textsc{sgn-arrakis}~\cite{sgnarrakis_repo} provides a source element for the \textsc{arrakis} time-series distribution service~\cite{arrakis} that is used in IGWN low-latency deployments.

\subsection{Pipeline topology}\label{sec:architecture:topology}

A single \textsc{sgnax} invocation processes an arbitrary list of channels from a single interferometer.
The pipeline is constructed by the analysis entry-point script (\texttt{sgnax-extract}, \texttt{sgnax-timeseries}, or \texttt{sgnax-qtransform}) according to the chosen data source and analysis algorithm; the package's remaining console entry points (\texttt{sgnax-dagger}, \texttt{sgnax-merge}, \texttt{sgnax-synchronize}, and \texttt{sgnax-archive}) implement the workflow layer of Sec.~\ref{sec:architecture:workflow} rather than pipelines of their own.
For each requested channel, the pipeline contains an independent dataflow chain originating at a source element and terminating at one or more event sinks.
Figure~\ref{fig:pipeline} shows the two resulting topologies.
For the matched-filter pipeline, the whitened time series of each channel fans out across the requested sampling rates, which are restricted to powers of two (Sec.~\ref{sec:algorithms:matched_filter}); each rate branch resamples to its rate, correlates against that rate's template bank with a \texttt{Correlate} element, and locates triggers with a \texttt{TorchPeakFinder}, after which a \texttt{WindowedTriggerAggregator} merges the per-rate trigger streams.
For the Q-transform pipeline, the whitened time series of each channel passes to a single \texttt{QScan} element that projects each chunk onto every requested Q-plane and applies cross-plane deduplication, after which a \texttt{TriggerClusterer} (enabled by default) merges tiles of the same glitch in time.
In both pipelines every channel's final trigger stream is routed to its own pad on a single shared \texttt{HDF5FeatureSink} (Sec.~\ref{sec:architecture:triggers}), which a \texttt{KafkaTriggerSink} replaces in online mode (Sec.~\ref{sec:architecture:workflow}).

\begin{figure*}[tp]
  \centering
  \resizebox{\textwidth}{!}{
\begin{tikzpicture}[
    font=\footnotesize,
    elt/.style={draw, rounded corners=1.8pt, align=center, line width=0.5pt,
                minimum height=9mm, minimum width=14mm, inner sep=2.6pt},
    src/.style={elt, fill=teal!14, draw=teal!55!black},
    tr/.style={elt, fill=blue!9, draw=blue!50!black},
    snk/.style={elt, fill=orange!22, draw=orange!75!black},
    ts/.style={-{Stealth[length=2mm,width=1.9mm]}, line width=0.65pt, draw=black!72},
    tsplain/.style={line width=0.65pt, draw=black!72},
    ev/.style={-{Stealth[length=2mm,width=1.9mm]}, line width=1.15pt,
               draw=violet!62!black},
    evplain/.style={line width=1.15pt, draw=violet!62!black},
    evopt/.style={-{Stealth[length=2mm,width=1.9mm]}, line width=1.15pt,
                  draw=violet!62!black, dashed},
    ratelbl/.style={font=\scriptsize\itshape, text=black!60},
    panellbl/.style={font=\normalsize\bfseries},
    note/.style={font=\scriptsize\itshape, text=black!58, align=center},
  ]

  \def\xSrc{0.0}  \def\xWh{2.7}
  \def\xRe{5.6}   \def\xAm{8.0}  \def\xCo{10.4}  \def\xPk{12.9}
  \def\xAg{15.9}  \def\xSk{18.6}
  \def\xFanO{4.1} \def\xFanI{14.5}
  \def\rT{0.0}  \def\rM{-1.6}  \def\rD{-2.7}  \def\rB{-3.8}  \def\rMid{-1.9}

  \node[panellbl,anchor=west] at (-0.8,1.2)
       {(a)\quad\texttt{sgnax-extract}\ \ ---\ \ multi-rate matched filter};

  \node[src] (aSrc) at (\xSrc,\rMid) {Data\\source};
  \node[tr]  (aWh)  at (\xWh,\rMid)  {Whiten};

  \foreach \y/\p in {\rT/T,\rM/M,\rB/B}{
    \node[tr] (Re\p) at (\xRe,\y) {Resampler};
    \node[tr] (Am\p) at (\xAm,\y) {Amplify};
    \node[tr] (Co\p) at (\xCo,\y) {Correlate};
    \node[tr] (Pk\p) at (\xPk,\y) {Torch\\PeakFinder};
  }
  \node[ratelbl] at (\xRe,0.78)  {rate $r_1$};
  \node[ratelbl] at (\xRe,-0.82) {rate $r_2$};
  \node[ratelbl] at (\xRe,-3.02) {rate $r_N$};
  \foreach \x in {\xRe,\xAm,\xCo,\xPk}{\node at (\x,\rD) {$\vdots$};}

  \node[tr] (aAg) at (\xAg,\rMid) {Windowed\\Trigger\\Aggregator};
  \node[snk] (aSk) at (\xSk,\rMid) {HDF5\\Feature\\Sink};

  \draw[ts] (aSrc) -- (aWh);
  \draw[tsplain] (aWh.east) -- (\xFanO,\rMid);
  \draw[tsplain] (\xFanO,\rT) -- (\xFanO,\rB);
  \foreach \y/\p in {\rT/T,\rM/M,\rB/B}{
    \draw[ts] (\xFanO,\y) -- (Re\p.west);
    \draw[ts] (Re\p) -- (Am\p);
    \draw[ts] (Am\p) -- (Co\p);
    \draw[ts] (Co\p) -- (Pk\p);
  }
  \foreach \y/\p in {\rT/T,\rM/M,\rB/B}{
    \draw[evplain] (Pk\p.east) -- (\xFanI,\y);
  }
  \draw[evplain] (\xFanI,\rT) -- (\xFanI,\rB);
  \draw[ev] (\xFanI,\rMid) -- (aAg.west);
  \draw[ev] (aAg) -- (aSk);

  \def\ybT{-5.5}    
  \def\ybB{-7.0}    
  \def\xQ{\xRe}     
  \def\xCluTr{8.4}  
  \def\xSkB{11.2}   
  \def\xFanB{6.9}   

  \node[panellbl,anchor=west] at (-0.8,-4.65)
       {(b)\quad\texttt{sgnax-qtransform}\ \ ---\ \ Q-transform excess power};

  \node[src] (bSrc) at (\xSrc,\ybT) {Data\\source};
  \node[tr]  (bWh)  at (\xWh,\ybT)  {Whiten};
  \node[tr]  (bQ)   at (\xQ,\ybT)   {QScan};
  \node[tr]  (bCl)  at (\xCluTr,\ybT) {Trigger\\Clusterer};
  \node[snk] (bSkC) at (\xSkB,\ybT) {HDF5\\Feature\\Sink};
  \node[snk] (bSkU) at (\xSkB,\ybB) {HDF5\\Feature\\Sink\\(unclustered)};

  \draw[ts] (bSrc) -- (bWh);
  \draw[ts] (bWh)  -- (bQ);
  \draw[evplain] (bQ.east) -- (\xFanB,\ybT);
  \draw[ev] (\xFanB,\ybT) -- (bCl.west);
  \draw[ev] (bCl) -- (bSkC);
  \draw[evopt] (\xFanB,\ybT) -- (\xFanB,\ybB) -- (bSkU.west);
  \node[note,anchor=east] at (\xFanB-0.25,\ybB)
       {\texttt{-{}-save-unclustered}\\(optional)};

  \def\xL{13.2}
  \node[src,minimum width=8mm,minimum height=5mm] at (\xL,\ybT) {};
  \node[anchor=west,font=\scriptsize] at (\xL+0.5,\ybT) {source};
  \node[tr,minimum width=8mm,minimum height=5mm] at (\xL+2.7,\ybT) {};
  \node[anchor=west,font=\scriptsize] at (\xL+3.2,\ybT) {transform};
  \node[snk,minimum width=8mm,minimum height=5mm] at (\xL+5.7,\ybT) {};
  \node[anchor=west,font=\scriptsize] at (\xL+6.2,\ybT) {sink};
  \draw[ts] (\xL-0.4,\ybB) -- (\xL+0.4,\ybB);
  \node[anchor=west,font=\scriptsize] at (\xL+0.5,\ybB)
       {time-series frame};
  \draw[ev] (\xL+3.5,\ybB) -- (\xL+4.3,\ybB);
  \node[anchor=west,font=\scriptsize] at (\xL+4.4,\ybB)
       {trigger (event) frame};

\end{tikzpicture}}
  \caption{%
    \textsc{sgn} dataflow topology of the two \textsc{sgnax} pipelines, for a single channel.
    (a) \texttt{sgnax-extract}: the whitened time series fans out across the requested power-of-two sampling rates; each rate branch resamples, applies the variance-restoring \texttt{Amplify} gain, correlates against that rate's template bank, and finds peaks, after which a \texttt{WindowedTriggerAggregator} merges the per-rate triggers into the channel's pad on the shared \texttt{HDF5FeatureSink} (under \texttt{-{}-save-rates} each branch's stream is additionally recorded, omitted here).
    (b) \texttt{sgnax-qtransform}: the whitened time series passes to a \texttt{QScan} element that performs the chunking, Q-plane projection, bisquare windowing, excess-power statistic, and cross-plane deduplication; a \texttt{TriggerClusterer} (\texttt{-{}-cluster-dt}, enabled by default) time-clusters its trigger stream before the sink, and \texttt{-{}-save-unclustered} optionally taps the raw \texttt{QScan} stream to a second sink of the same layout (dashed).
    Both pipelines are built from the same \texttt{DataSourceInfo}-configured source-and-whitening front end; only the processing downstream of whitening differs, and in online mode a \texttt{KafkaTriggerSink} replaces the file sink (Sec.~\ref{sec:architecture:workflow}).
    Arrow styles distinguish time-series frames from trigger (event) frames.%
  }
  \label{fig:pipeline}
\end{figure*}

The two pipelines share a configuration object, \texttt{DataSourceInfo}, that is constructed from the command-line arguments and that fully specifies the data acquisition stage.
The same channel list, time range, and source-selection arguments produce a topologically identical input subgraph for both algorithms; only the per-channel processing chain downstream of whitening differs.
This factoring lets us add data-source backends in one place and have both analyses pick them up.

\subsection{Data sources}\label{sec:architecture:sources}

The \texttt{DataSourceInfo} object accepts one of three backends, listed in Table~\ref{tab:datasources}.
All three backends feed the same on-the-fly whitening stage before either analysis; whitening is enabled by default and can be disabled by a command-line flag for debugging.
The \texttt{frames} backend reads from a frame-cache file (\texttt{.lcf}) listing GWF frames.
The \texttt{devshm} backend reads from the shared-memory ring buffer used for low-latency operation at the IGWN sites; it is selected for online deployments where new frames appear in \texttt{/dev/shm} every second.
The \texttt{arrakis} backend reads from the \textsc{arrakis} distribution service~\cite{arrakis}, a new low-latency data-delivery service currently deployed at the LIGO sites and intended to serve IGWN-wide distribution in future observing runs.
Taken together, the \texttt{devshm} and \textsc{arrakis} backends bring \texttt{sgnax-qtransform} into low-latency operation, streaming Q-transform triggers to detector commissioners within tens of seconds of acquisition, well ahead of the cadence at which existing production \textsc{omicron} deployments deliver triggers in practice.
A complete description of each backend, its configuration parameters, and its supported analysis modes is given in Appendix~\ref{sec:appendix:data_sources}.

\begin{table}[t]
  \caption{%
    Data sources supported by \textsc{sgnax}.
    The third column indicates whether the source is appropriate for offline (O) or low-latency online (L) operation.
    All sources share the same whitening stage, enabled by default; \texttt{devshm} denotes the backend reading the \texttt{/dev/shm} path.%
  }
  \label{tab:datasources}
  \begin{tabular}{lll}
    \hline\hline
    Source & Description & Mode \\
    \hline
    \texttt{frames}   & GWF cache via \textsc{sgn-ligo} & O \\
    \texttt{devshm}   & \texttt{/dev/shm} ring buffer               & L \\
    \texttt{arrakis}  & \textsc{arrakis} timeseries service~\cite{arrakis} & L \\
    \hline\hline
  \end{tabular}
\end{table}

\subsection{Trigger output schema}\label{sec:architecture:triggers}

Triggers from all pipelines are written to \textsc{hdf5} by the \texttt{HDF5FeatureSink} element, in a layout readable by \textsc{gwpy}'s \texttt{hdf5.snax} reader: one group per channel, containing one structured dataset per \texttt{-{}-cadence} of analyzed time whose columns are the trigger fields \texttt{timestamp}, \texttt{time}, \texttt{snr}, \texttt{phase}, \texttt{frequency}, \texttt{q}, and \texttt{duration}.
A dataset is written for every configured channel each epoch, even when a channel produced no triggers, so coverage gaps are explicit in the file rather than inferred from absence.
Datasets are appended incrementally as each cadence of data completes rather than buffered to end-of-stream, so that interrupted runs leave a partial file in a consistent state; in online operation the sink rotates output files on a configurable persist cadence.
Under \texttt{-{}-save-rates} the per-rate, per-channel matched-filter streams are additionally recorded as per-rate groups in the same file with the same schema, allowing the same downstream consumers to read both raw and aggregated triggers; for streaming deployments a \texttt{KafkaTriggerSink} publishes the same records to \textsc{kafka} topics (Sec.~\ref{sec:architecture:workflow}).

The schema is intentionally narrower than the \textsc{omicron} \textsc{root} \texttt{TTree} format~\cite{robinet2020omicron, root1997}, which carries additional fields used by the \textsc{omicron} HTML reporting tooling.
The \textsc{python}-native format is chosen for compatibility with the \textsc{hdf5} readers in the broader detector-characterization toolchain.
A future revision is planned to add an \textsc{omicron}-compatible writer for downstream tools that have not yet been migrated.

\subsection{Workflow generation and deployment}\label{sec:architecture:workflow}

Production analyses are deployed through a single DAG generator, \texttt{sgnax-dagger}, which builds \textsc{htcondor} workflows.
Offline runs divide the requested observing time among parallel jobs and merge their per-channel outputs.
Online runs launch long-lived searches that publish triggers to \textsc{kafka}; synchronization and archival stages merge these streams in time and write rotating feature files.
The same analysis configuration serves both deployments, replacing the separate workflow tooling of the predecessors.
Configuration formats, frame transfers, and packaging are described in Appendix~\ref{sec:appendix:implementation:workflow}.

\subsection{Comparison with the predecessor architectures}\label{sec:architecture:comparison}

The architectural differences from \textsc{snax} and \textsc{omicron} can be summarized as follows.
Relative to \textsc{snax}, \textsc{sgnax} replaces the \textsc{gstreamer}-1.0 dataflow~\cite{cannon2021gstlal} with the in-process \textsc{sgn} graph and removes the corresponding C-level dependencies; multi-channel coverage is expressed within the pipeline graph rather than only at the workflow level, with each \textsc{htcondor} job analyzing many channels through a single graph; and the \textsc{kafka}-based streaming layer of \textsc{snax} is reproduced by the \textsc{kafka} trigger sink and source together with the synchronize and archive stages of Sec.~\ref{sec:architecture:workflow}.
Relative to \textsc{omicron}, \textsc{sgnax} replaces a C++ executable with a \textsc{python}-importable package, removes the \textsc{root} and \textsc{gwollum} dependencies, replaces the bespoke parameter-file format with command-line arguments that can equivalently be supplied through a \textsc{yaml} configuration file, and writes triggers to \textsc{hdf5} rather than \textsc{root} \texttt{TTree} files.
The \textsc{omicron} algorithm itself is reproduced in Sec.~\ref{sec:algorithms:qtransform} below, with the same bisquare windowing, log-spaced tiling, and excess-power statistic.
The package dependencies, source-tree layout, and the static-analysis and test infrastructure of \textsc{sgnax} are described in Appendix~\ref{sec:appendix:implementation}.

\section{Algorithms}\label{sec:algorithms}
\subsection{Multi-rate sine-Gaussian matched filter}\label{sec:algorithms:matched_filter}

The \texttt{sgnax-extract} entry point implements the same statistic as \textsc{snax}~\cite{snax_repo}: a sliding-window matched-filter correlation against a bank of unit-norm sine-Gaussian templates, Eqs.~\eqref{eq:matched_filter}--\eqref{eq:snr_iq}.
The bank is populated by a mismatch-driven log-spaced grid in $(f, Q)$ derived from the sine-Gaussian metric of Chatterji~\cite{chatterji2005}; we reproduce the spacing rules in Appendix~\ref{sec:appendix:bank_spacing}.
At a maximum mismatch $\mu_{\max}$, the bank places templates such that the match between any signal in the parameter space and its nearest template is at least $1 - \mu_{\max}$.

\begin{figure*}[tp]
  \centering
  \includegraphics[width=0.96\textwidth]{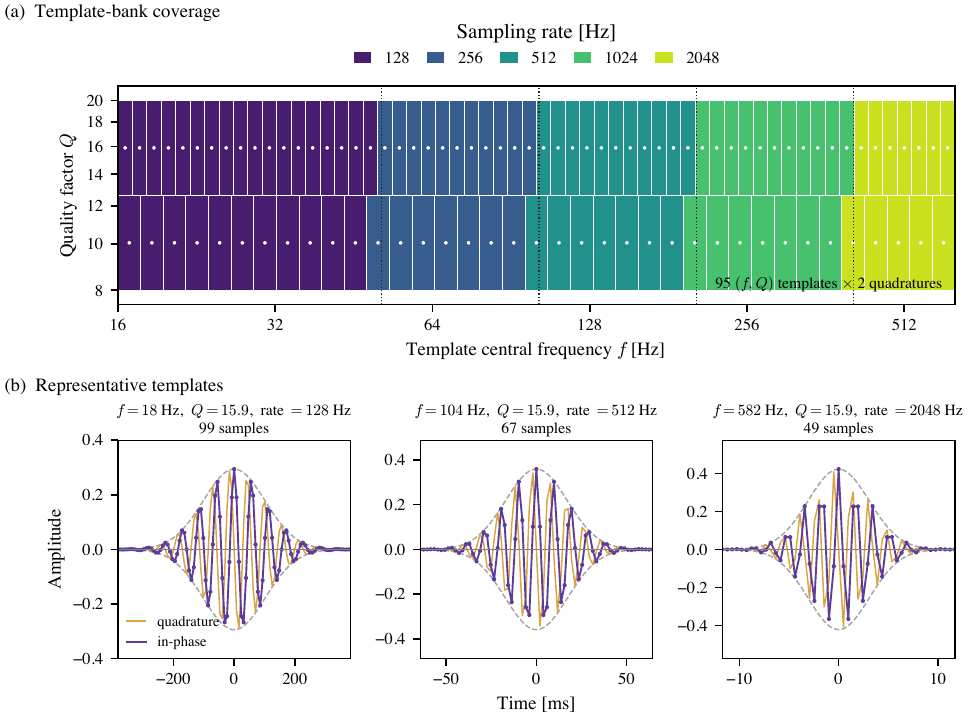}
  \caption{%
    The \texttt{sgnax-extract} sine-Gaussian matched-filter template bank, for the default configuration ($f \in [20, 800]\,\mathrm{Hz}$, $Q \in [8, 20]$, maximum mismatch $\mu_{\max} = 0.2$, sampling rates $\{128, 256, 512, 1024, 2048\}\,\mathrm{Hz}$).
    (a) Coverage of the $(f, Q)$ parameter space: each rectangle is the region for which one template is the closest match, colored by the power-of-two sampling rate the template is assigned to, with dotted lines marking the rate breakpoints $\alpha\, r_k / 2$.
    The higher-$Q$ row carries a denser frequency grid, following Eq.~\eqref{eq:nf}.
    (b) Representative sine-Gaussian templates (in-phase and quadrature, with the Gaussian envelope dashed), each evaluated at its assigned rate; every template resolves to a few tens of samples because the rate tracks the template frequency.%
  }
  \label{fig:template_bank}
\end{figure*}

The bank is partitioned across a user-specified set of power-of-two sampling rates $\{r_k\}$; the restriction to powers of two is a requirement of the \textsc{sgn-ts} \texttt{Resampler}, which produces each rate branch from the whitened working rate by successive factor-of-two decimations.
A template with central frequency $f$ is assigned to the smallest rate $r_k$ such that $f \leq \alpha\, r_k / 2$, with $\alpha < 1$ (default $0.8$) chosen to keep templates away from the Nyquist anti-aliasing rolloff.
Templates assigned to rate $r_k$ are zero-padded to a common length and normalized so that all bank vectors have unit $\ell^2$-norm at the rate at which they are evaluated.
Whitened input data are resampled into each requested rate; the multi-rate fan-out is implemented as a set of parallel branches in the \textsc{sgn} graph, each terminated by an independent \texttt{TorchPeakFinder} and trigger sink.
A final \texttt{WindowedTriggerAggregator} merges across rates by retaining the highest-SNR trigger per fixed-width time window, eliminating duplicate detections of a single transient by templates of similar morphology at adjacent rates.
Figure~\ref{fig:template_bank} shows the resulting bank for the default configuration, with panel~(a) giving the $(f, Q)$ coverage colored by assigned sampling rate and panel~(b) representative templates.

The correlation of Eq.~\eqref{eq:matched_filter} is evaluated by the \texttt{Correlate} element of \textsc{sgn-ts}~\cite{huang2026sgn}, which evaluates the discrete cross-correlation only where template and data fully overlap.
Templates are stored as \texttt{torch.Tensor} objects in single precision; the correlation executes on the device selected by the \textsc{pytorch} backend (CPU by default, GPU when one is available)~\cite{paszke2019pytorch}.
The two quadrature outputs of each $(f, Q)$ template pair are combined via Eq.~\eqref{eq:snr_iq} inside the \texttt{TorchPeakFinder}, which then identifies, within each fixed-duration window, the template pair with the highest combined SNR and emits a single trigger record per window.
The trigger phase is recovered from the in-phase and quadrature components as $\varphi = \arctan2(\rho_{f,Q,\pi/2}, \rho_{f,Q,0})$.
When multiple channels share a sampling rate, the per-channel resampling, correlation, and peak-finding elements can further be replaced by batched counterparts (\texttt{-{}-batch-channels}) that evaluate all such channels in single \textsc{pytorch} operations, reducing both the element count of the graph and the per-channel dispatch overhead on CPU and GPU alike.

The principal differences relative to \textsc{snax} are: (i) the multi-rate bank construction has been moved into a single \texttt{SineGaussianBank} class in \textsc{sgnax}, replacing a set of \textsc{python} helpers in the \textsc{snax} waveforms module; (ii) the correlation runs on the \textsc{pytorch} backend rather than through \textsc{gstlal}'s C-level filter elements; (iii) cross-rate aggregation is performed by a single dataflow element rather than by post-processing of \textsc{hdf5} files; and (iv) all channels requested in a single invocation are processed within one pipeline graph, with concurrency exposed at the dataflow-element level, so that a production run covers a large channel list by distributing it across multiple \textsc{htcondor} jobs, each analyzing many channels within one graph.
The statistic itself, including the per-rate mismatch grid and the quadrature SNR estimator, is unchanged.

\paragraph{Timeseries feature mode.}
The production role of \textsc{snax} is not the sparse trigger list of the preceding paragraphs but a dense, regularly-sampled \emph{feature timeseries}: at a fixed cadence (typically $16\,\mathrm{Hz}$), the loudest $(\rho, f, Q, \varphi)$ feature in each time bin is recorded for every channel regardless of its SNR, producing the fixed-shape feature vectors consumed by machine-learning inference frameworks such as \textsc{idq}~\cite{essick2021idq, godwin2020thesis}.
The \texttt{sgnax-timeseries} entry point reproduces this mode with the identical multi-rate correlation front end: the per-rate peak finders run with no SNR threshold, a windowed aggregator merges the per-rate streams---closing each window only once every rate branch has delivered data past it, so that the differing filter latencies of the rate branches cannot exclude a lagging branch's contribution---and a resampling stage places the surviving features onto an absolute GPS-aligned grid, emitting exactly one feature per bin.
The output is dense and gap-free by construction ($\sim\!1.4\times10^{6}$ rows per channel-day at $16\,\mathrm{Hz}$), with empty bins recorded explicitly, and is written to the same \textsc{hdf5} schema as the trigger files.

\subsection{Q-transform excess-power search}\label{sec:algorithms:qtransform}

The \texttt{sgnax-qtransform} entry point reimplements the \textsc{omicron} algorithm~\cite{robinet2020omicron, robinet2018technote} as a single \textsc{sgn} transform element, \texttt{QScan}.
The whitened time series is partitioned into chunks of duration $T_{\rm chunk}$ (default $64\,\mathrm{s}$) with overlap $T_{\rm ovl}$, and each chunk is multiplied by a Tukey window with shape parameter $\alpha = T_{\rm ovl} / T_{\rm chunk}$ before being transformed by an in-place real-to-complex FFT.

\begin{figure*}[tp]
  \centering
  \includegraphics[width=0.96\textwidth]{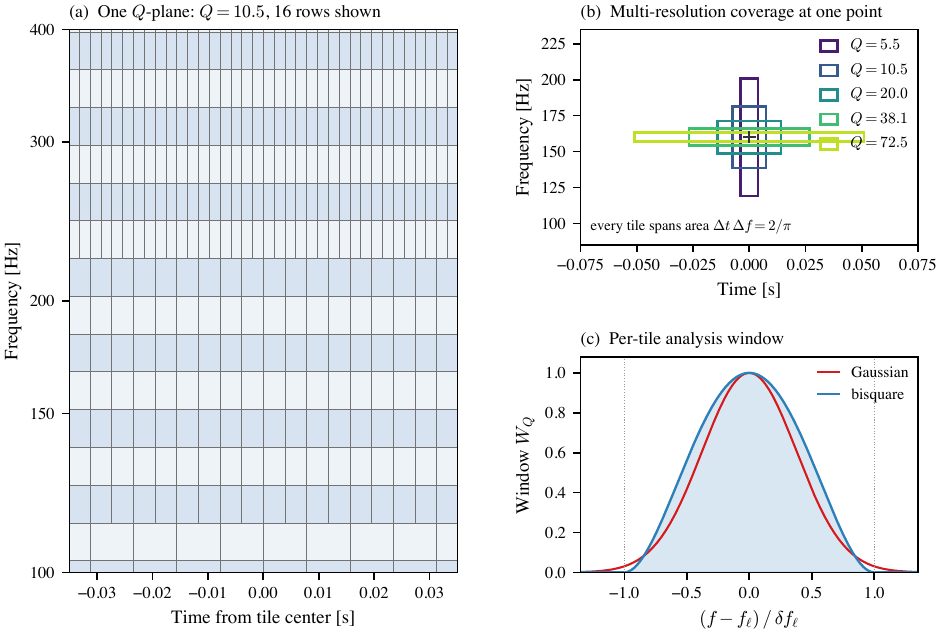}
  \caption{%
    The \texttt{sgnax-qtransform} time--frequency tiling, for the default configuration ($Q \in [4, 100]$, $f \in [20, 2048]\,\mathrm{Hz}$, maximum mismatch $\mu_{\max} = 0.2$, $64\,\mathrm{s}$ chunk).
    (a) The tiling of a single Q-plane: logarithmically-spaced frequency rows, each subdivided into linearly-spaced time tiles whose width narrows with frequency.
    (b) The tile that each of the five Q-planes places over a common time--frequency point; the tiles share a fixed area $\Delta t\,\Delta f = 2/\pi$ but trade time resolution for frequency resolution as $Q$ increases.
    (c) The bisquare analysis window of Eq.~\eqref{eq:bisquare} (shaded), compared with a Gaussian of the same second moment---the window of the original constant-Q transform~\cite{brown1991qtransform}, which the finitely supported bisquare approximates.%
  }
  \label{fig:qtiling}
\end{figure*}

For each requested $Q$, the algorithm constructs a Q-plane: a logarithmic grid of central frequencies $\{f_\ell\}$ spanning the user-specified range, with bandwidths $\delta f_\ell = f_\ell\sqrt{11}/Q$.\footnote{The upper edge of the frequency range is bounded by the whitening Nyquist frequency, half the working sample rate; spanning to the $2048\,\mathrm{Hz}$ of the configuration shown in Fig.~\ref{fig:qtiling} therefore requires a working rate of at least $4096\,\mathrm{Hz}$.}
Each frequency band is endowed with a bisquare window of unit-variance normalization,
\begin{equation}\label{eq:bisquare}
  W_Q(f; f_\ell)
    = \mathcal{N}_Q
      \left[1 - \left(\frac{f - f_\ell}{\delta f_\ell}\right)^{\!2}\right]^2,
  \qquad |f - f_\ell| \leq \delta f_\ell,
\end{equation}
with $\mathcal{N}_Q$ chosen so that $\mathbb{E}\bigl[\lvert X^{\rm wh}\rvert^2\bigr] = 2$ for unit-variance whitened Gaussian noise, matching the \textsc{omicron} convention~\cite[Sec.~2]{robinet2020omicron}.
Because triggers are only retained from the un-tapered interior of the Tukey-windowed chunk (the edge trim below), the per-bin input variance entering $\mathcal{N}_Q$ is that of the un-windowed chunk, $N = T_{\rm chunk} f_s$, rather than the chunk-averaged value $\sum_k w_k^2 = N(1 - 5\alpha/8)$ of the Tukey window.
The two choices differ by a uniform SNR scale factor $(1 - 5\alpha/8)^{-1/2}$---$2.0\%$ at the default $\alpha = 1/16$---to which the near-threshold trigger counts of the production comparison in Sec.~\ref{sec:results:omicron} are directly sensitive.
The complex transform coefficient at the band center is recovered by a centered inverse FFT in which the $f_\ell$ bin maps to the IFFT zero-frequency bin.
Figure~\ref{fig:qtiling} illustrates the construction: the time--frequency grid of a single Q-plane (panel a), the multi-resolution coverage provided by the full set of Q-planes (panel b), and the bisquare window of Eq.~\eqref{eq:bisquare} (panel c).

For each tile $(\tau, f_\ell, Q)$, the excess-power signal-to-noise ratio~\cite{anderson2001excesspower} is
\begin{equation}\label{eq:sgnax_qtransform_snr}
  \hat\rho^2(\tau, f_\ell, Q)
    = \max\!\bigl(\lvert X^{\rm wh}\rvert^2 - 2,\; 0\bigr),
\end{equation}
identical to Eq.~\eqref{eq:omicron_snr}~\cite[Eq.~2]{robinet2020omicron}.
Tiles whose SNR exceeds a configured threshold are retained as candidate triggers.
The algorithm trims a region of width $T_{\rm ovl}/2$ from each chunk edge to suppress Tukey-tapered artifacts before applying a cross-plane deduplication step that keeps the loudest tile within a small time window across all $Q$-planes.
An optional time-clustering pass merges triggers within a configurable separation $\Delta t_{\rm cl}$ via greedy single-linkage clustering, retaining the loudest tile's SNR and representative parameters while extending the duration over the cluster.
The per-tile SNR cut is applied before cross-plane deduplication and time clustering; the default cluster statistic follows the loudest-tile convention of \textsc{omicron}, rather than summing the power of overlapping tiles.

The Q-plane spacing follows the standard mismatch metric for the constant-Q transform~\cite{chatterji2004multiresolution, chatterji2005}, also recapitulated in Appendix~\ref{sec:appendix:bank_spacing}.
The number of Q-planes is set by
\begin{equation}\label{eq:num_q_planes}
  N_Q = \left\lceil
    \frac{1}{2\sqrt{\mu_{\max}/3}}\,
    \frac{1}{\sqrt{2}}\,
    \ln\!\left(\frac{Q_{\max}}{Q_{\min}}\right)
  \right\rceil,
\end{equation}
and the number of frequency bands within each Q-plane by an analogous expression with the Q-dependent metric coefficient.
At fixed $\mu_{\max} = 0.2$, $Q \in [\sqrt{11}, 100]$, $f \in [20, 2048]\,\mathrm{Hz}$, and a $64\,\mathrm{s}$ chunk, the resulting tile count agrees with the \textsc{omicron} configuration of Robinet \emph{et al.}~\cite{robinet2020omicron} to within rounding.

The principal differences relative to \textsc{omicron} are implementation-level rather than algorithmic.
\textsc{sgnax}'s \texttt{QScan} runs in-process under \textsc{python}, vectorizes per-band IFFTs over all bands within a Q-plane via \textsc{numpy} array operations, and integrates with the same data-source backends as the matched-filter pipeline.
Despite its pure-\textsc{python} execution, \texttt{QScan} is competitive with \textsc{omicron}'s C++ core~\cite{robinet2020omicron, omicron_repo} in single-channel throughput---in fact modestly faster on the benchmark of Sec.~\ref{sec:results:runtime}---and amortizes further as channels are added (Sec.~\ref{sec:results:scaling}).
\textsc{sgnax} targets the single-interferometer glitch identification on which this paper is focused; cross-detector coincidence is by design outside its scope.

\paragraph{Device execution.}
The projection of each chunk onto the $Q$-planes is optionally executed on a GPU (\texttt{-{}-device}).
The chunking, Tukey window, and forward real-to-complex FFT remain host-side double-precision operations, preserving the numerics of the default path; the per-band scatter of spectral coefficients, the batched complex inverse FFTs over the bands of each plane, and the tile threshold scan then run on the device in single-precision complex arithmetic through \textsc{pytorch}~\cite{paszke2019pytorch}, using index tensors precomputed at configuration time and shared across identically configured channels, with only the sparse above-threshold tiles transferred back to the host.
With \texttt{-{}-batch-channels} the whitening front end is likewise batched, whitening the channels of each sampling-rate group through single batched double-precision FFT operations---on the CPU by default, or on the GPU for CUDA devices.
Trigger output is unchanged to single-precision rounding ($\sim\!10^{-7}$ relative in SNR); the capacity this execution mode provides on production hardware is measured in Sec.~\ref{sec:results:gpu}.

\section{Results}\label{sec:results}

We validate each of the two \textsc{sgnax} analyses by injecting a population of sine-Gaussian waveforms into a stationary Gaussian-noise auxiliary channel and recovering them with \texttt{sgnax-extract} and \texttt{sgnax-qtransform} in turn.
The injection set consists of $N_{\rm inj} = 3299$ sine-Gaussian waveforms drawn over the parameter ranges of Table~\ref{tab:run_config}; injection times are uniform over a $T_{\rm obs} = 10^{5}\,\mathrm{s}$ segment of a synthetic LIGO Hanford (H1) auxiliary channel, and the recovered trigger lists are matched to the injection table within a coincidence window of $\Delta t = 0.25\,\mathrm{s}$, retaining the loudest trigger per injection.
The injection set is generated, injected into the synthetic channel, and matched to the recovered triggers by the Glitch Validation Toolkit (\textsc{gvt})~\cite{gvt_repo}; the same matching procedure is used for both pipelines.

\begin{table}[t]
  \caption{%
    Parameters of the injection-recovery validation run reported in this section.
    The auxiliary channel is a synthetic unit-variance Gaussian time series at the LIGO Hanford (H1) nominal sampling rate; the injection set spans the full parameter range covered by the \textsc{sgnax} bank.%
  }
  \label{tab:run_config}
  \begin{tabular}{ll}
    \hline\hline
    Parameter & Value \\
    \hline
    Channel                                       & \texttt{H1:FAKE-STRAIN} \\
    Observation time $T_{\rm obs}$                & $10^{5}\,\mathrm{s}$ \\
    Injections $N_{\rm inj}$                      & $3299$ \\
    Frequency range $[f_{\min}, f_{\max}]$        & $[40, 3000]\,\mathrm{Hz}$ \\
    Quality factor $[Q_{\min}, Q_{\max}]$         & $[3, 9]$ \\
    Injected SNR range                            & $[2.9, 48.5]$ \\
    Bank mismatch $\mu_{\max}$                    & $0.05$ \\
    SNR threshold                                 & $5.5$ \\
    Coincidence window $\Delta t$                 & $0.25\,\mathrm{s}$ \\
    \hline\hline
  \end{tabular}
\end{table}

\subsection{Matched-filter injection recovery}\label{sec:results:extract}

\begin{figure*}[tp]
  \centering
  \includegraphics[width=\textwidth]{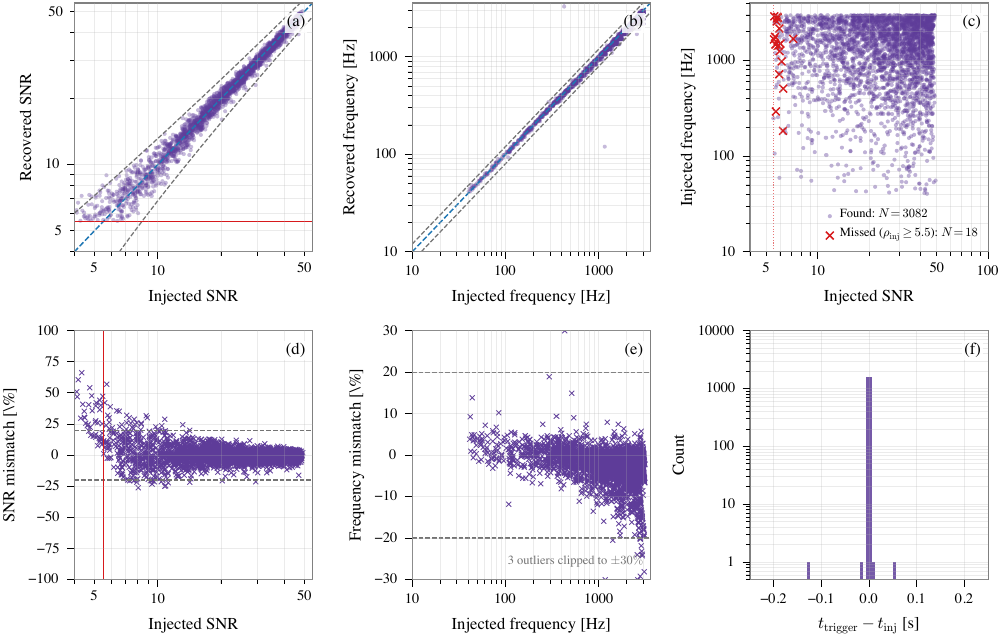}
  \caption{%
    Injection recovery for the \texttt{sgnax-extract} matched-filter pipeline on the validation set of Table~\ref{tab:run_config}.
    (a)~Recovered vs. injected SNR; the blue dashed line is the identity, the gray dashed lines mark the expected $\rho_{\rm inj} \pm \sqrt{\rho_{\rm inj}}$ scatter band, and the red line marks the SNR threshold.
    (b)~Recovered vs. injected central frequency; bands are $\pm 20\%$.
    (c)~Found and above-threshold missed injections in the (injected SNR, injected frequency) plane; injections drawn with $\rho_{\rm inj} < 5.5$ are unrecoverable by construction and are omitted.
    (d)~Fractional SNR error against injected SNR.
    (e)~Fractional frequency error against injected frequency.
    (f)~Trigger-time residual relative to the injection time, on a log count axis.%
  }
  \label{fig:validation_extract}
\end{figure*}

Figure~\ref{fig:validation_extract} summarizes the injection-recovery performance of \texttt{sgnax-extract}.
The injection set spans $\rho_{\rm inj} \in [2.9, 48.5]$, so $153$ of the $3299$ injections are drawn below the analysis threshold of $\rho = 5.5$ and are not in principle recoverable.
Of the $3146$ recoverable injections, $3128$ ($99.4\%$) are matched within the coincidence window; of the remaining $18$ above-threshold misses, all but one were injected with $\rho_{\rm inj} < 6.3$ (the last at $\rho_{\rm inj} = 7.2$), and they lie predominantly at frequencies above $\sim 1\,\mathrm{kHz}$ (panel c, red), as expected for a sine-Gaussian matched filter operating against a bank of finite mismatch.
The $153$ sub-threshold injections are not algorithmic misses but a consequence of the analysis threshold itself; they are omitted from panel~(c).
The recovered SNR (panel a) and central frequency (panel b) cluster on the diagonal; $99.6\%$ of the matched injections recover SNR within the expected $\pm\sqrt{\rho_{\rm inj}}$ scatter band (dashed), modulo a mild positive bias near the SNR threshold consistent with the noise floor of the matched-filter statistic.
The SNR mismatch (panel d) is contained within the $\pm 20\%$ band for $\rho_{\rm inj} \gtrsim 8$, with the larger scatter at low SNR being the expected near-threshold behavior.
The fractional frequency error (panel e) is contained within $\pm 20\%$ across the full bank, with sub-percent residual bias at intermediate frequencies and a $\sim 3\%$ downward bias near the upper edge of the bank where templates approach the Nyquist anti-aliasing rolloff.
The time-residual distribution (panel f) is sharply peaked at zero with a $\sim 1\,\mathrm{ms}$ characteristic width, indicating that the multi-rate fan-out of Sec.~\ref{sec:algorithms:matched_filter} preserves trigger time to the resolution of the highest rate.

\subsection{Q-transform injection recovery}\label{sec:results:qtransform}

\begin{figure*}[tp]
  \centering
  \includegraphics[width=\textwidth]{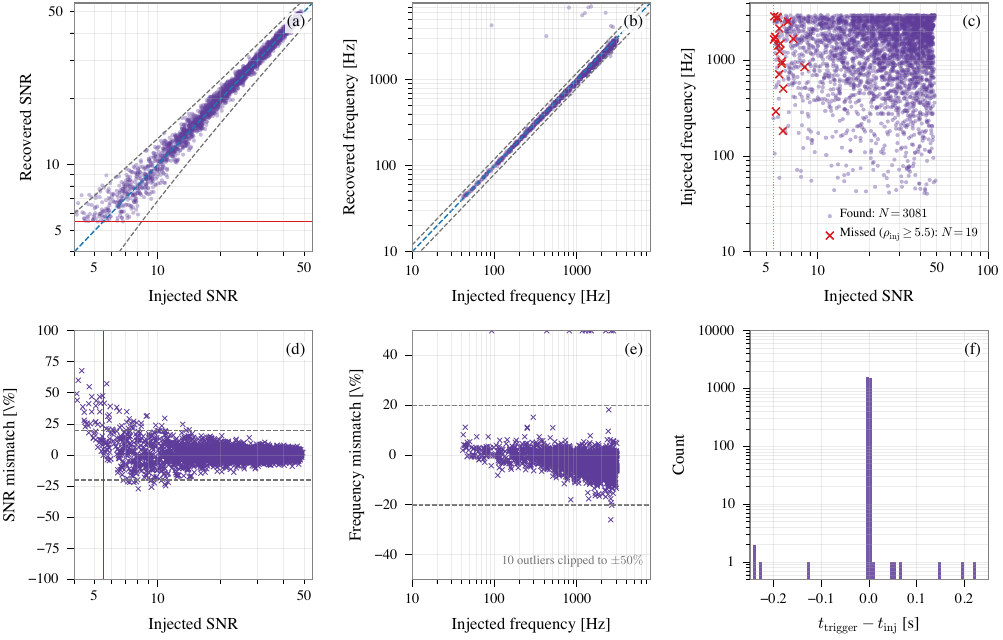}
  \caption{%
    Injection recovery for the \texttt{sgnax-qtransform} Q-transform pipeline on the same validation set as Fig.~\ref{fig:validation_extract}.
    Panel content and conventions follow Fig.~\ref{fig:validation_extract}.
    Frequency-mismatch outliers exceeding the panel limit (e) are clipped and tallied in the annotation; they correspond to injections recovered in an adjacent Q-plane near the bank edge.%
  }
  \label{fig:validation_qtransform}
\end{figure*}

Figure~\ref{fig:validation_qtransform} reports the analogous injection-recovery test for \texttt{sgnax-qtransform}.
\texttt{sgnax-qtransform} matches $3130$ of the $3299$ injections within the coincidence window ($99.4\%$ of the recoverable population, on par with the matched-filter pipeline); this count includes three of the $153$ injections drawn below $\rho = 5.5$, which upward noise fluctuations carry above threshold.
The $19$ above-threshold misses (panel c, red) are concentrated near the SNR threshold and at the upper end of the analysis band, consistent with the constant-Q tile structure becoming sparse at the bank edges; the remaining $150$ sub-threshold injections are unrecoverable by construction and are omitted from panel~(c).
SNR and frequency recovery (panels a and b) lie on the diagonal across more than two decades.
The SNR mismatch (panel d) closely tracks the matched-filter case, contained within $\pm 20\%$ for $\rho_{\rm inj} \gtrsim 8$ and broadening at low SNR, reflecting the more stochastic nature of the excess-power statistic near threshold.
The frequency mismatch (panel e) is contained within $\pm 20\%$ for the vast majority of injections; a small population of outliers occurs near the Q-plane boundaries and is consistent with the deduplication choice of Sec.~\ref{sec:algorithms:qtransform}.
The time-residual distribution (panel f) shows a tight central peak with somewhat broader wings than the matched-filter pipeline, consistent with the chunked Tukey-windowed FFT used by the Q-transform.

\subsection{Comparison of the two analysis modes}\label{sec:results:modes}

\begin{figure*}[tp]
  \centering
  \includegraphics[width=\textwidth]{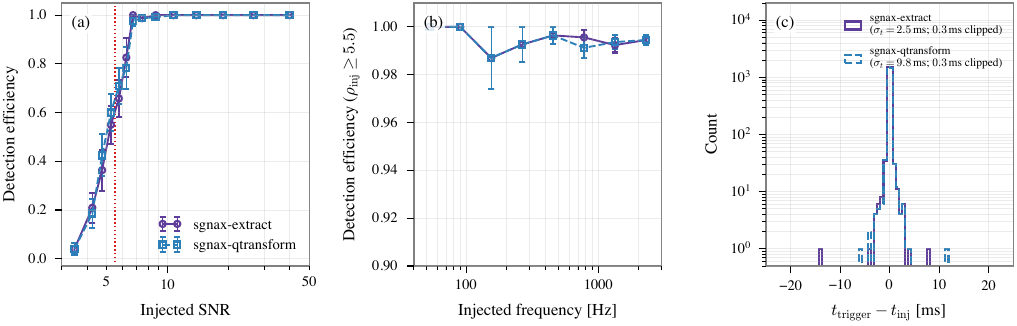}
  \caption{%
    Head-to-head comparison of the two \textsc{sgnax} analysis modes on the shared injection set of Table~\ref{tab:run_config}.
    (a)~Detection efficiency versus injected SNR, with binomial uncertainties; the dotted red line marks the $\rho = 5.5$ threshold.
    (b)~Detection efficiency versus injected frequency for the recoverable ($\rho_{\rm inj} \geq 5.5$) injections.
    (c)~Trigger-time residual relative to the injection time, on a log count axis, with the per-mode standard deviation $\sigma_t$ annotated both for the full distribution and with the isolated $|\Delta t| > 10\,\mathrm{ms}$ outliers clipped.
    The two modes are equivalent detectors (panels a, b) and localize triggers with the same core precision; the larger unclipped $\sigma_t$ of the chunked Q-transform is driven by a handful of far outliers (panel c).%
  }
  \label{fig:mode_comparison}
\end{figure*}

Because the matched-filter and Q-transform pipelines are run on the same injection set, their recovery can be compared directly; Figure~\ref{fig:mode_comparison} does so along three axes.
As detectors of the injected population the two modes are equivalent: their detection efficiency turns on at the SNR threshold in lockstep (panel a) and is flat and consistent with unity across the full frequency band for recoverable injections (panel b), so the choice between them is not a question of detection power.
The two modes also recover injected SNR and central frequency with statistically indistinguishable accuracy (Secs.~\ref{sec:results:extract}--\ref{sec:results:qtransform}).
Time localization (panel c) separates the modes only in the far tails of the residual distribution: the central peaks are equally tight---clipping the handful of isolated outliers beyond $\pm 10\,\mathrm{ms}$ leaves $\sigma_t = 0.34\,\mathrm{ms}$ for both modes---but the Q-transform produces roughly four times as many such outliers ($11$ versus $3$ of the $\approx 3100$ matched injections), which dominate its unclipped width ($\sigma_t \approx 9.8\,\mathrm{ms}$ versus $2.5\,\mathrm{ms}$) and are consistent with occasional localization to an adjacent tile in its coarser, frequency-dependent time tiling.

The more consequential difference between the two modes is architectural and concerns trigger latency.
The Q-transform applies an FFT to each analysis chunk and so cannot emit a trigger until the chunk enclosing that time has been buffered in full, fixing its trigger latency at order the chunk duration $T_{\rm chunk}$ (Sec.~\ref{sec:algorithms:qtransform}; tens of seconds in production configurations).
The matched filter of \texttt{sgnax-extract}, by contrast, is a streaming correlation that carries no per-chunk buffering: its algorithmic lag is only the template support, at most tens of milliseconds, so it can in principle generate triggers inline with the incoming channel data.
In practice its end-to-end latency is dominated not by the trigger algorithm but by the shared whitening and acquisition front end, which we measure at a few seconds (Sec.~\ref{sec:results:latency}); the Q-transform adds its chunk duration on top of this floor.
This makes \texttt{sgnax-extract} fast enough to serve as a source of low-latency auxiliary-channel features for tools that must consume detector-characterization data on the cadence of the frames, such as the \textsc{idq} statistical noise-inference framework~\cite{essick2021idq, Huxford:2024vdt}, whereas the per-chunk FFT of \texttt{sgnax-qtransform} sets a floor of tens of seconds on its trigger latency at the default chunk---ample for low-latency commissioning scans, but not for feature generation on the cadence of the frames themselves.
Run on the same archival strain with its bank extended to the full $10\,\mathrm{Hz}$--$8\,\mathrm{kHz}$ range, the matched filter recovers a closely matched low-frequency trigger population and the same loud events as the Q-transform, differing mainly in a denser near-threshold high-frequency background; we develop this strain-level comparison in Sec.~\ref{sec:results:omicron}.

\subsection{Broadband white-noise-burst recovery}\label{sec:results:wnb}

The injection sets above are sine-Gaussians, the localized waveforms for which a single template or tile is, by construction, a good model.
A more demanding test---and the one on which the event-trigger-generator studies of McIver~\cite{mciver2015thesis} and Godwin~\cite{godwin2020thesis} stressed the recovery of bulk parameters---is the band-limited white-noise burst (BTLWNB), whose power is spread across a finite bandwidth rather than concentrated at a single point in the time--frequency plane.
We inject $1200$ BTLWNB waveforms (central frequency $100$--$550\,\mathrm{Hz}$, bandwidth $50$--$500\,\mathrm{Hz}$, duration $10$--$100\,\mathrm{ms}$, injected SNR uniform over $[5, 48]$) into a unit-variance white-noise channel.
In white noise the matched-filter optimal SNR of an injection equals the discrete $\ell^2$ norm of its samples, so the injected SNR is fixed exactly with no reference-PSD dependence.
A population of $100$ sine-Gaussians injected alongside is recovered to within a few percent in SNR (median $\rho_{\rm rec}/\rho_{\rm inj} = 0.95$--$0.96$ for both modes), confirming the calibration.

\begin{figure*}[tp]
  \centering
  \includegraphics[width=0.82\textwidth]{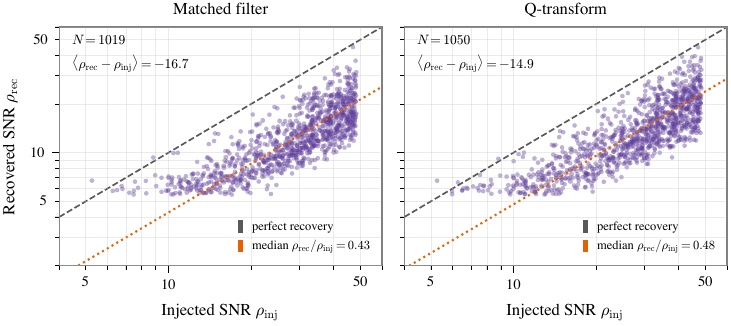}
  \caption{%
    Recovered versus injected SNR for $1200$ band-limited white-noise-burst injections, for the matched-filter (left) and Q-transform (right) modes, on log--log axes.
    The dashed line is perfect recovery; the dotted line is the median recovered-to-injected SNR ratio, annotated together with the found-injection count $N$ and the mean SNR offset $\langle\rho_{\rm rec}-\rho_{\rm inj}\rangle$.
    Both modes under-recover the SNR of these broadband events---a single template or tile captures only part of the spread power---by amounts consistent with the production trigger generators benchmarked in Ref.~\cite{mciver2015thesis}, with the Q-transform's loudest-tile statistic recovering marginally more than the single-template matched filter.%
  }
  \label{fig:wnb_recovery}
\end{figure*}

Figure~\ref{fig:wnb_recovery} shows the recovery.
Both modes detect the broadband population efficiently---$85\%$ (\texttt{sgnax-extract}) and $88\%$ (\texttt{sgnax-qtransform}) of the recoverable injections---but both systematically underestimate the total event SNR, recovering a median of $0.43$ (matched filter) and $0.48$ (Q-transform) of the injected value, for mean offsets of $-16.7$ and $-14.9$ respectively.
This is the expected behavior of a single-template or single-tile loudness estimate applied to a non-localized event: a single sine-Gaussian template, or a single $Q$-plane tile, captures only the fraction of a broadband burst's power that falls within its own time--frequency support~\cite{mciver2015thesis}.
The offsets reproduce the values McIver tabulated for the production event-trigger generators of the time on the same waveform class (\textsc{omicron}, mean $-13.0$; DMT-Omega, $-15.6$)~\cite{mciver2015thesis}.
The Q-transform recovers marginally more of the broadband power than the matched filter, but both default statistics describe the loudest template or tile rather than the total power of the event.
Accurate total-SNR recovery for broadband bursts was identified as an open problem in~\cite{mciver2015thesis}; we show next that the configurability of \textsc{sgnax} already lets a user recover substantially more of this power than the default single-tile statistic.

\begin{figure}[tp]
  \centering
  \includegraphics[width=\columnwidth]{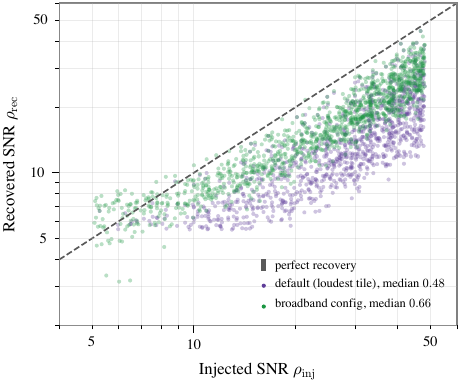}
  \caption{%
    A configurable broadband-SNR estimator for \texttt{sgnax-qtransform}, built from the pipeline's unclustered tile dump.
    Recovered versus injected SNR for the band-limited white-noise bursts of Fig.~\ref{fig:wnb_recovery}, under the default single-loudest-tile statistic (median $0.48$) and the non-overlapping cluster-SNR that sums the excess power of an event's mutually non-overlapping tiles within the best-matched $Q$-plane (median $0.66$ at a per-tile threshold of $3.0$).
    Lowering the per-tile threshold retains more of the spread sub-threshold power and recovers a larger fraction of the injected SNR, but is bounded below perfect recovery by the threshold itself; the localized sine-Gaussian recovery is unchanged by this estimator, so the configuration trades broadband completeness against tile count without biasing the localized population.%
  }
  \label{fig:wnb_broadband}
\end{figure}

\paragraph{A configurable broadband estimator.}
The default per-tile statistic is the right choice for the localized transients that dominate detector-characterization work, but the same scriptable configuration that makes \textsc{sgnax} easy to deploy also lets a user trade it off deliberately for broadband events.
Because \texttt{sgnax-qtransform} can emit its full unclustered tile dump (\texttt{-{}-save-unclustered}) and exposes the per-tile threshold (\texttt{-{}-snr-threshold}), a dedicated broadband configuration can sum the excess power across an event's spread tiles rather than reporting the single loudest.
Summing naively over the overcomplete tiling over-counts power---a median recovered-to-injected ratio of $1.4$ summing within a $Q$-plane and $2.4$ summing across all planes, reproducing the caution of Ref.~\cite{mciver2015thesis}---but restricting the sum to mutually non-overlapping tiles within the best-matched $Q$-plane avoids the double counting.
This non-overlapping cluster-SNR recovers a median of $0.62$ of the injected broadband SNR at the default threshold, against $0.48$ for the single loudest tile, while leaving the localized sine-Gaussian recovery unchanged at $0.96$ (Fig.~\ref{fig:wnb_broadband}).
Lowering the per-tile threshold extends the gain---to $0.66$ at a threshold of $3.0$---as more of the sub-threshold spread power is retained, at the cost of a rapidly growing tile count; the residual deficit is set by this threshold, below which the spread power never enters the trigger set.
A group concerned specifically with broadband noise can therefore recover such events more completely than the single-tile statistic of the predecessor tools allows by standing up a dedicated \textsc{sgnax} process with these settings, a one-line change to the shared configuration rather than a separate pipeline.

Beyond loudness, the time and frequency of the broadband injections are recovered without bias.
Table~\ref{tab:wnb_recovery} collects the recovery statistics: both modes localize the BTLWNB events in time to a standard deviation of about $22\,\mathrm{ms}$ and in central frequency to about $70\,\mathrm{Hz}$, against $0.6\,\mathrm{ms}$ and $20$--$45\,\mathrm{Hz}$ for the localized sine-Gaussians injected alongside.
The broader broadband residuals are expected---a white-noise burst's power is spread in both time and frequency, so the loudest tile or template need not sit at the injected centroid---but they carry no systematic offset, and they are tighter than the corresponding figures McIver tabulated for the production event-trigger generators (\textsc{omicron} frequency standard deviation $199\,\mathrm{Hz}$ on this waveform class)~\cite{mciver2015thesis}.

\begin{table}[t]
  \caption{%
    Time and frequency recovery for the band-limited white-noise-burst (BTLWNB) and sine-Gaussian (SG) injections of Sec.~\ref{sec:results:wnb}, for the matched-filter (\texttt{sgnax-extract}) and Q-transform (\texttt{sgnax-qtransform}) modes.
    Entries are the mean and standard deviation of the recovered-minus-injected time and central frequency over the found injections of each class; the broadband events carry larger residuals than the localized sine-Gaussians but no systematic bias.%
  }
  \label{tab:wnb_recovery}
  \begin{tabular}{llccccc}
    \hline\hline
    Mode & Class & $N$ & $\langle\Delta t\rangle$ & $\sigma_t$ & $\langle\Delta f\rangle$ & $\sigma_f$ \\
         &       &     & [ms] & [ms] & [Hz] & [Hz] \\
    \hline
    \texttt{extract}    & BTLWNB & $1019$ & $-0.4$ & $23.0$ & $-8.5$  & $64.8$ \\
                        & SG     & $98$   & $\phantom{-}0.0$  & $\phantom{0}0.6$  & $-19.4$ & $45.0$ \\
    \texttt{qtransform} & BTLWNB & $1050$ & $\phantom{-}0.2$  & $22.2$ & $\phantom{-}1.6$   & $75.2$ \\
                        & SG     & $97$   & $-0.1$ & $\phantom{0}0.6$  & $-5.0$  & $22.4$ \\
    \hline\hline
  \end{tabular}
\end{table}

\subsection{Detection efficiency and false-alarm rate}\label{sec:results:roc}

\begin{figure}[tp]
  \centering
  \includegraphics[width=0.92\columnwidth]{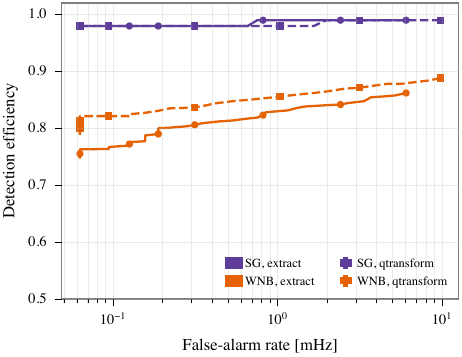}
  \caption{%
    Detection efficiency versus false-alarm rate for the two \textsc{sgnax} modes (\texttt{extract}, solid; \texttt{qtransform}, dashed) and the two waveform classes (sine-Gaussian, purple; band-limited white-noise burst, orange), traced by varying the SNR threshold.
    Efficiency is measured from the injection MDC of Sec.~\ref{sec:results:wnb}; the false-alarm rate from a separate $32{,}000\,\mathrm{s}$ pure-white-noise run.
    The broadband curves lie below the sine-Gaussian curves at every false-alarm rate, the ROC counterpart of the broadband SNR under-recovery of Fig.~\ref{fig:wnb_recovery}.%
  }
  \label{fig:roc}
\end{figure}

The injection studies above quantify detection efficiency, but efficiency is only meaningful against the rate of false alarms it is bought with.
Following the event-trigger-generator validation of McIver~\cite{mciver2015thesis}, we trace the two together as the SNR threshold is varied (Fig.~\ref{fig:roc}).
Efficiency is measured from the injection MDC of Sec.~\ref{sec:results:wnb}, separately for the sine-Gaussian and broadband populations; the false-alarm rate is measured from a separate $32{,}000\,\mathrm{s}$ run on pure white noise, as the rate of triggers with no injection within the coincidence window.
Both modes recover sine-Gaussians with $\sim\!0.98$ efficiency all the way down to a sub-mHz false-alarm rate, and the false-alarm rate itself falls steeply with threshold---from $\sim\!10\,\mathrm{mHz}$ at $\rho = 5.5$ to below $0.1\,\mathrm{mHz}$ by $\rho = 6.5$, with no false alarm at all above $\rho \approx 7$ in $32{,}000\,\mathrm{s}$ for either mode---confirming the clean separation between the signal and Gaussian-noise trigger populations.
The broadband curves sit below the sine-Gaussian curves at every false-alarm rate, by a margin (efficiency $\sim\!0.85$ versus $\sim\!0.98$ at $1\,\mathrm{mHz}$) that is the direct ROC counterpart of the SNR under-recovery of Sec.~\ref{sec:results:wnb}: because a broadband event's loudness is underestimated, fewer such events clear a given SNR threshold even though their near-threshold detection efficiency is high.

\subsection{\textsc{omicron} comparison on production data}\label{sec:results:omicron}

The production references are chosen to match the available data products: \textsc{omicron} strain triggers and \textsc{snax} auxiliary-channel feature streams.
Both \textsc{sgnax} algorithms are also compared on the same strain channel below and on the shared synthetic data of Sec.~\ref{sec:results:modes}; their input is not restricted to the channel type of either production reference.
The matched-filter strain runs in this paper are validation studies, rather than an established production use; the intended strain trigger product is the Q-transform output.
The injection-recovery validation of Sec.~\ref{sec:results:qtransform} establishes that \texttt{sgnax-qtransform} recovers the input parameters of synthetic sine-Gaussian transients to within the expected mismatch.
To check that this carries over to real detector data, we run \texttt{sgnax-qtransform} on the LIGO Livingston strain channel \texttt{L1:GDS\_CALIB\_STRAIN\_NOLINES} for a 24-hour interval beginning at 2025-01-01 00:00:00 UTC (GPS $1419724818$) and compare the resulting trigger list directly with the \textsc{omicron} production triggers for the same channel and interval.

\begin{figure*}[tp]
  \centering
  \includegraphics[width=\textwidth]{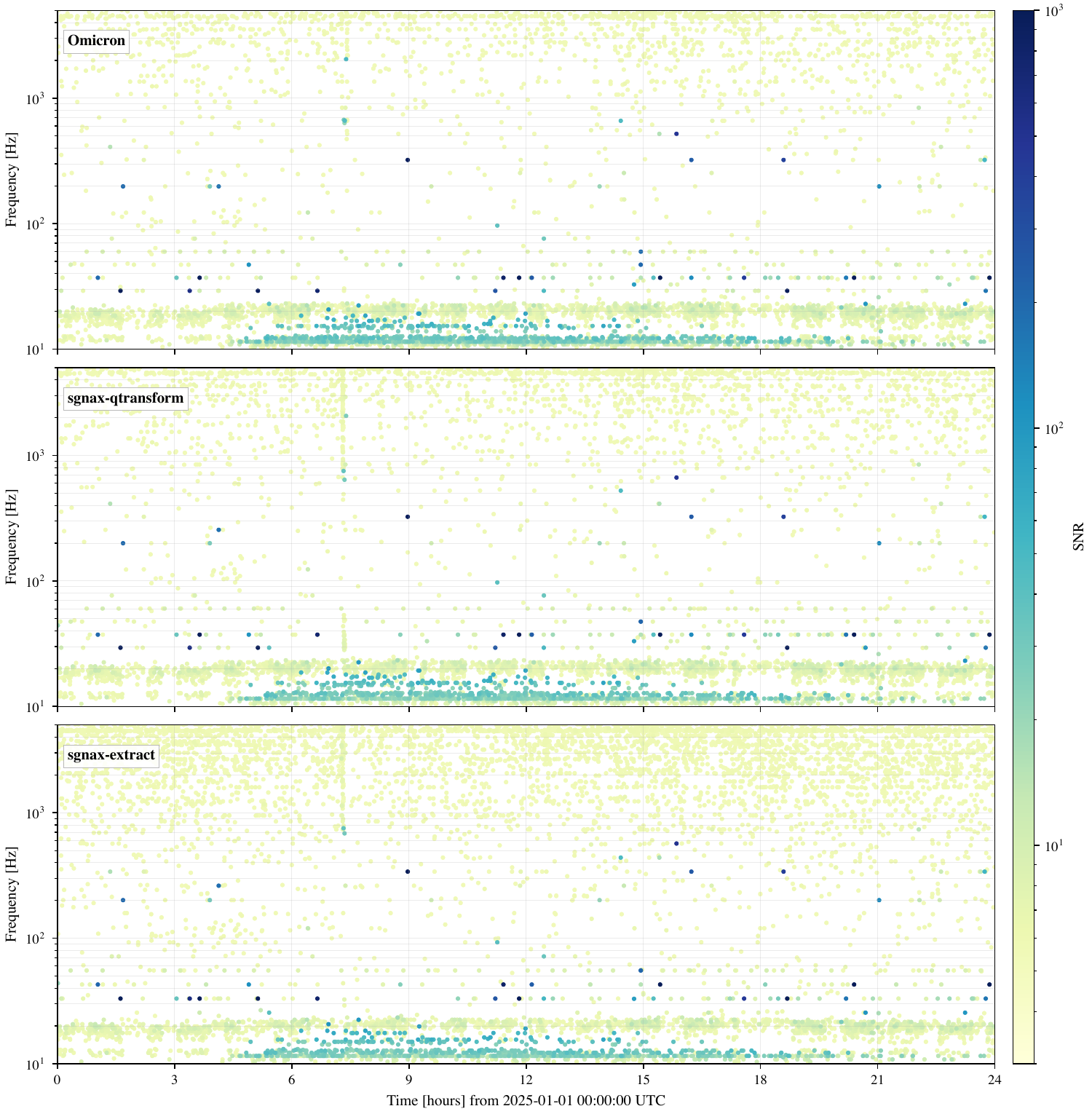}
  \caption{%
    Twenty-four hours of triggers on \texttt{L1:GDS\_CALIB\_STRAIN\_NOLINES} beginning 2025-01-01 00:00:00 UTC (GPS $1419724818$), in the time--frequency plane and colored by SNR on a common logarithmic scale.
    Top: production \textsc{omicron} triggers~\cite{robinet2020omicron}.
    Middle: \texttt{sgnax-qtransform} triggers on the same channel and interval, matching the production analysis's four $Q$-planes, per-tile SNR threshold, and clustering window.
    Bottom: \texttt{sgnax-extract} matched-filter triggers, with the multi-rate bank extended to span $10\,\mathrm{Hz}$--$8\,\mathrm{kHz}$ and greedy time-clustered ($0.1\,\mathrm{s}$ window) to match the production clustering transform.
    The top two panels report the same excess-power statistic and are directly comparable; \texttt{sgnax-extract} reports matched-filter SNR, so for the bottom panel the shared color scale is a morphological guide rather than a quantitative SNR match.%
  }
  \label{fig:omicron_vs_sgnax}
\end{figure*}

The top two panels of Figure~\ref{fig:omicron_vs_sgnax} show the \textsc{omicron} and \texttt{sgnax-qtransform} trigger populations in the time--frequency plane, colored by SNR on a common logarithmic scale.
The \texttt{sgnax-qtransform} run matches the \textsc{omicron} production analysis in every discrete configuration choice---the $Q$ range and maximum mismatch (hence the same four $Q$-planes), the per-tile SNR threshold of $5.5$, and the $0.1\,\mathrm{s}$ clustering window---with the whitening PSD estimated over an $8\,\mathrm{s}$ FFT.
The two trigger lists agree in fine structure across the plane: both recover the persistent band of sub-$30\,\mathrm{Hz}$ excursions, the narrow-band line populations between roughly $30$ and $200\,\mathrm{Hz}$, and the cluster of high-SNR transients between the sixth and twelfth hours of the run, including the prominent column of broadband triggers near hour $7.5$; the five loudest events of the day coincide in time and frequency between the two pipelines.
The full-band trigger totals agree to $0.1\%$ ($11{,}683$ for \textsc{omicron} against $11{,}670$ for \texttt{sgnax-qtransform}), with the sub-$2\,\mathrm{kHz}$ counts agreeing to $5\%$ ($5.5 \times 10^{3}$ versus $5.3 \times 10^{3}$) and \texttt{sgnax-qtransform} returning $4\%$ more triggers above $2\,\mathrm{kHz}$ ($6.4 \times 10^{3}$ versus $6.1 \times 10^{3}$), where the paired per-trigger SNRs of the two pipelines agree to $0.4\%$ in the median.

This level of agreement is a sensitive test of the SNR calibration, because the near-threshold trigger distribution is steep: on this interval the count above threshold falls by a factor of $\sim 2$ per half-unit of SNR, so a fractional SNR-scale offset $\epsilon$ shifts the count above a fixed threshold by roughly $9\epsilon$.
The variance-scale convention in the tile normalization $\mathcal{N}_Q$ of Eq.~\eqref{eq:bisquare} illustrates the stakes: substituting the chunk-averaged per-bin variance of the Tukey-windowed input for the un-windowed value appropriate to the trimmed chunk interior (Sec.~\ref{sec:algorithms:qtransform}) would inflate every reported SNR by $2.0\%$ and appear here as a $\sim 20\%$ count shift concentrated in the near-threshold tiles.
Trigger counts near a threshold thus amplify percent-level calibration differences by roughly an order of magnitude; the paired per-trigger SNR ratio is the sharper diagnostic, and we quote both.

A remaining population-level difference lies in the loud, low-frequency triggers: \textsc{omicron} records $821$ triggers above $\mathrm{SNR} = 20$ against $732$ for \texttt{sgnax-qtransform}, with the difference concentrated below $30\,\mathrm{Hz}$ during the sustained scattered-light period spanning roughly hours $5$--$17$ of the run.
Trigger-by-trigger pairing localizes the mechanism: during that period the paired \texttt{sgnax-qtransform} SNRs sit $\sim 4\%$ below \textsc{omicron}'s, while outside it the two agree at the percent level.
The pipelines differ in how the whitening normalization adapts to sustained noise: \textsc{omicron} whitens each analysis chunk twice, with the second pass dividing by the measured spectrum of the once-whitened chunk, so its SNR scale renormalizes to the elevated local noise floor; the \textsc{sgn-ligo} whitener used by \textsc{sgnax} tracks a median-based spectral estimate over a $\sim 2$-minute timescale that holds closer to the stationary floor, the convention required for a matched-filter-consistent SNR scale~\cite{cannon2021gstlal}.
Neither convention is uniquely correct---one stabilizes the effective single-tile false-alarm rate through noise transients, the other preserves a stable physical normalization---and the difference is confined to periods of sustained non-stationarity.
The present comparison establishes that the $Q$-plane construction and excess-power statistic of Sec.~\ref{sec:algorithms:qtransform}, validated on synthetic injections above, reproduce the \textsc{omicron} trigger population on archival strain data to $0.1\%$ in total count, with per-trigger SNRs matched at the percent level and the residual differences localized to this documented difference in whitening adaptation.

We also run \texttt{sgnax-extract} on the same interval, both to exhibit the matched-filter mode on real strain and because the result bears on how the two modes divide the analysis (bottom panel of Fig.~\ref{fig:omicron_vs_sgnax}).
The multi-rate bank is extended upward to span the same $10\,\mathrm{Hz}$--$8\,\mathrm{kHz}$ range; because the matched filter emits a trigger at every template threshold crossing, we collapse its $1.5\times10^{5}$ raw triggers with the same greedy single-linkage time clustering used in production (a $0.1\,\mathrm{s}$ window, with each cluster reported using its loudest tile's parameters), giving $1.3\times10^{4}$ clustered events spanning $8\,\mathrm{Hz}$ to $6.5\,\mathrm{kHz}$.
The two modes report different detection statistics---matched-filter SNR for \texttt{sgnax-extract}, excess power for \texttt{sgnax-qtransform}---so the comparison of interest is where in the time--frequency plane each places triggers.
Below $2\,\mathrm{kHz}$ the two recover a closely matched population: $5.0\times10^{3}$ clustered \texttt{sgnax-extract} triggers against the $5.3\times10^{3}$ of \texttt{sgnax-qtransform}, with the loud population essentially identical---both show the persistent sub-$30\,\mathrm{Hz}$ excursion band and the same high-SNR isolated events, and all $617$ \texttt{sgnax-extract} triggers above SNR $20$ fall below $2\,\mathrm{kHz}$.
In the $10$--$100\,\mathrm{Hz}$, $\mathrm{SNR}>8$ band of greatest detector-characterization relevance (Sec.~\ref{sec:results:comparison}) the matched filter records $1.5\times10^{3}$ triggers, consistent with the \textsc{omicron} and \texttt{sgnax-qtransform} counts there.

Above $2\,\mathrm{kHz}$ the two modes diverge in count but not in loud content: \texttt{sgnax-extract} returns $7.7\times10^{3}$ clustered triggers, about $20\%$ more than the $6.4\times10^{3}$ of \texttt{sgnax-qtransform}, but this excess is almost entirely near-threshold---its median SNR is $5.7$ and fewer than ten of the high-frequency triggers exceed SNR $8$---the expected imprint of comparing a densely gridded matched-filter statistic against the coarser $Q$-plane tiling near a fixed threshold (Sec.~\ref{sec:results:omicron}).
Spanning the full band with the matched filter is, moreover, not free: each added sample rate enlarges the template bank and lengthens the runtime relative to the low-band configuration of Sec.~\ref{sec:results:runtime}, whereas the Q-transform covers the same range at fixed per-chunk cost.
The matched filter's distinctive value is thus concentrated where its streaming, low-latency output and dense low-frequency template spacing are decisive---the scattered-light and microseismically driven noise of the $10$--$100\,\mathrm{Hz}$ band (Sec.~\ref{sec:results:modes})---while the Q-transform is the more economical route to clean, uniform full-band coverage; this is the division of labor for which \textsc{sgnax} retains both modes.

\subsection{Comparison with predecessor pipelines}\label{sec:results:comparison}

Table~\ref{tab:residuals} collects per-statistic residual metrics against both predecessors---trigger-rate ratios, SNR-distribution Kolmogorov--Smirnov statistics, median per-trigger SNR differences, and per-trigger time offsets---since each isolates a distinct aspect of the predecessor reproduction discussed in Sec.~\ref{sec:algorithms}.
The \textsc{omicron} column is fixed by the strain-channel comparison of Sec.~\ref{sec:results:omicron}; the \textsc{snax} column is measured in Sec.~\ref{sec:results:snax} on the timeseries feature mode against the \textsc{snax} production features for the same 24-hour interval.

\begin{table}[t]
  \caption{%
    Residual differences between \textsc{sgnax} and its predecessors, both measured on the 24-hour interval beginning 2025-01-01 00:00:00 UTC.
    The \textsc{omicron} column is measured on the \texttt{L1:GDS\_CALIB\_STRAIN\_NOLINES} comparison over the full $10\,\mathrm{Hz}$--$8\,\mathrm{kHz}$ band, with no selection beyond the shared per-tile threshold; the time entry is the mean over paired tiles.
    The \textsc{snax} column is measured on the dense $16\,\mathrm{Hz}$ feature streams of four L1 auxiliary channels (Sec.~\ref{sec:results:snax}), restricted to the common $12$--$100\,\mathrm{Hz}$ template band; the rate and KS entries use a fixed $\mathrm{SNR} > 8$ selection on both pipelines, the SNR and time entries are medians over coincident loud features, and all four residuals are dominated by the multiband amplitude error of the production deployment localized in the text.%
  }
  \label{tab:residuals}
  \begin{tabular}{lcc}
    \hline\hline
    Statistic & vs.\ \textsc{snax} & vs.\ \textsc{omicron} \\
    \hline
    Trigger-rate ratio                   & $0.46$ & $1.00$ \\
    SNR-distribution KS statistic        & $0.49$ & $0.04$ \\
    Median per-trigger SNR difference    & $-30\%$ & $-0.3\%$ \\
    Per-trigger time difference [ms]     & $-8$ & $+0.3$ \\
    \hline\hline
  \end{tabular}
\end{table}

\begin{figure*}[tp]
  \centering
  \includegraphics[width=0.98\textwidth]{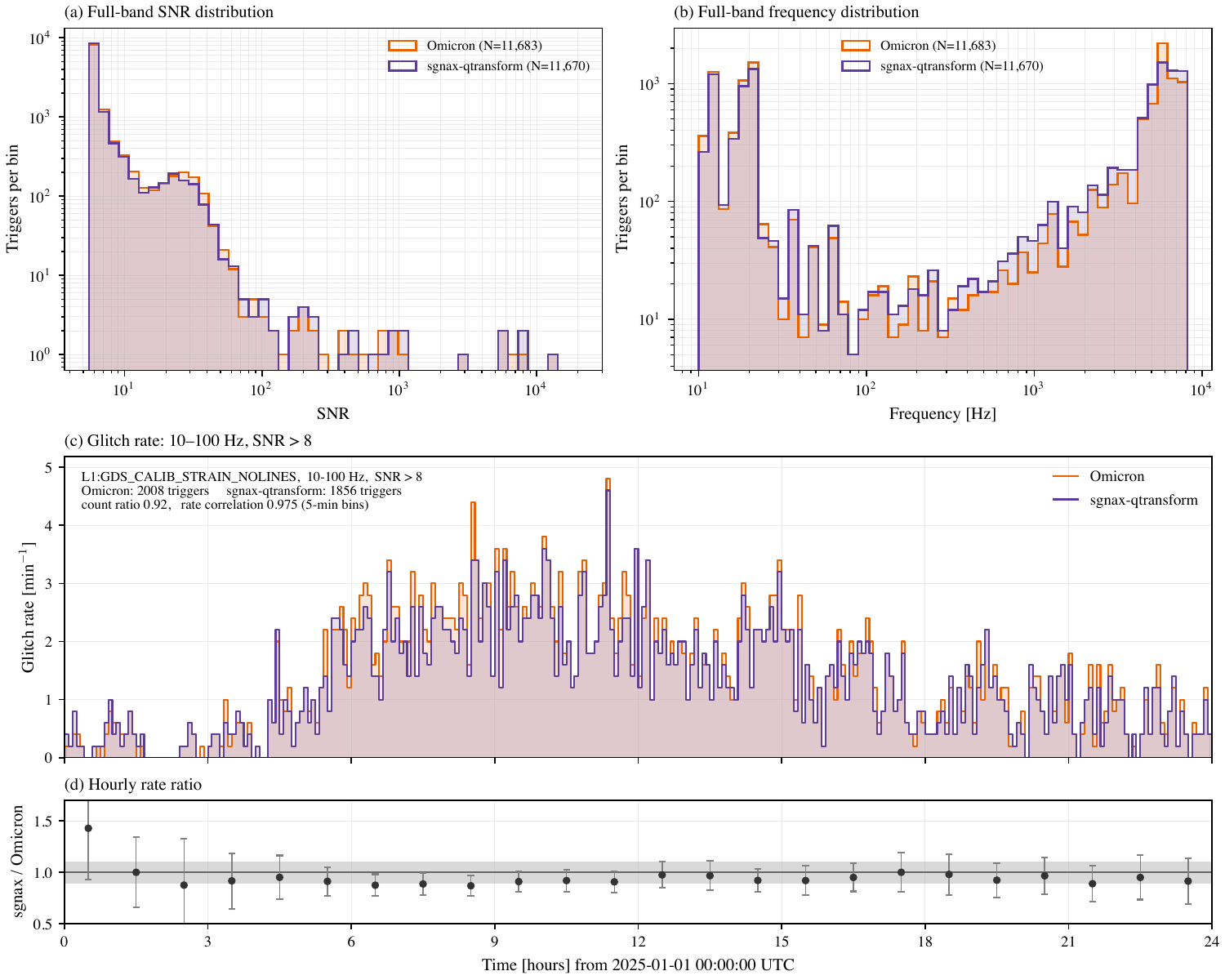}
  \caption{%
    Quantitative comparison of the \textsc{omicron} and \texttt{sgnax-qtransform} glitch populations on \texttt{L1:GDS\_CALIB\_STRAIN\_NOLINES} for the 24-hour interval beginning 2025-01-01 00:00:00 UTC.
    (a)~Full-band SNR distributions and (b)~frequency distributions, shown as trigger counts in common logarithmic bins, with no selection beyond the shared per-tile SNR threshold of $5.5$ and no normalization of the population totals ($11{,}683$ and $11{,}670$).
    (c)~Per-minute glitch rate in $5$-minute bins for the $10$--$100\,\mathrm{Hz}$, $\mathrm{SNR} > 8$ population.
    (d)~The corresponding \texttt{sgnax-qtransform}-to-\textsc{omicron} rate ratio in one-hour bins with Poisson uncertainties; the shaded band marks $\pm 10\%$ of unity.
    In this restricted population the pipelines record $2008$ and $1856$ triggers (rate ratio $0.92$), and their $5$-minute binned rate series correlate at $0.975$.%
  }
  \label{fig:glitch_rate}
\end{figure*}

The \textsc{omicron} column of Table~\ref{tab:residuals} is filled from the strain-channel comparison over the full $10\,\mathrm{Hz}$--$8\,\mathrm{kHz}$ band, with no selection beyond the shared per-tile threshold.
Each \textsc{omicron} trigger is paired with the nearest \texttt{sgnax-qtransform} trigger falling within a $0.2\,\mathrm{s}$ coincidence window and a factor-of-two frequency tolerance, in a one-to-one assignment that resolves the closest pairs first.
This pairs $69\%$ of the \textsc{omicron} triggers---the unpaired remainder is concentrated near threshold, where differences in preprocessing and candidate selection can change which tiles cross it---yielding a median per-trigger SNR difference of $-0.3\%$ and a mean per-tile time difference of $+0.3\,\mathrm{ms}$ consistent with zero, while the two full-band SNR distributions return a Kolmogorov--Smirnov statistic of $0.04$.
The agreement in total count does not imply identical event recovery: $31\%$ of the \textsc{omicron} triggers remain unpaired under this criterion, and the count differences below and above $2\,\mathrm{kHz}$ partly cancel.
The agreement requires no band selection: the median paired SNR difference is within $\pm 3\%$ in every frequency band, bounded by the $-3.1\%$ of the $10$--$30\,\mathrm{Hz}$ band in which the whitening-adaptation difference of Sec.~\ref{sec:results:omicron} is concentrated.
Figure~\ref{fig:glitch_rate} compares the glitch populations quantitatively: panels (a) and (b) show the full-band SNR and frequency distributions in common logarithmic bins, retaining the differences in absolute counts.
Panels (c) and (d) examine the time dependence of the $10$--$100\,\mathrm{Hz}$, $\mathrm{SNR} > 8$ population, where scattered light and several other transient-noise classes central to detector-characterization work are concentrated~\cite{davis2021detchar}: \textsc{omicron} records $2008$ triggers and \texttt{sgnax-qtransform} $1856$ (rate ratio $0.92$), the $5$-minute binned rate series correlate at $0.975$, and the hourly ratio remains within $\pm 10\%$ of unity across three quarters of the run.
For the \textsc{omicron} column, the shared per-tile threshold is applied before clustering and the table counts the final reported events, without a further SNR cut.
For the \textsc{snax} column, the $\mathrm{SNR} > 8$ cut selects bins of the dense feature stream; these are feature-bin counts rather than time-clustered event counts.
The \textsc{snax} column of Table~\ref{tab:residuals} is measured next.

\subsection{\textsc{snax} comparison on production auxiliary channels}\label{sec:results:snax}

The \textsc{snax} column of Table~\ref{tab:residuals} is measured on the timeseries feature mode of Sec.~\ref{sec:algorithms:matched_filter}, the dense regularly-sampled output that constitutes \textsc{snax}'s production data product.
We run \texttt{sgnax-timeseries} on the same 24-hour interval as the strain-channel comparisons above (2025-01-01, GPS $1419724818$) for the four L1 auxiliary channels---spanning seismic-isolation and length-sensing subsystems at native rates from $1024$ to $16384\,\mathrm{Hz}$---that are also present in the \textsc{snax} production feature archive for that day, and compare the two $16\,\mathrm{Hz}$ feature streams row by row.
The production analysis ran \textsc{snax}'s tapered-sine-Gaussian bank over roughly $12$--$1600\,\mathrm{Hz}$; the \texttt{sgnax-timeseries} configuration matches each channel's band from above, extends the template floor to $8\,\mathrm{Hz}$, and adopts the template density of the production deployment itself---effective mismatch $0.03$ with $Q \in [\sqrt{11}, 40]$, reproducing the nine-value production $Q$ ladder exactly---so that the comparison is template-for-template; the residual statistics are restricted to the common $12$--$100\,\mathrm{Hz}$ band in which the loud features of these channels concentrate.

\begin{figure*}[tp]
  \centering
  \includegraphics[width=0.98\textwidth]{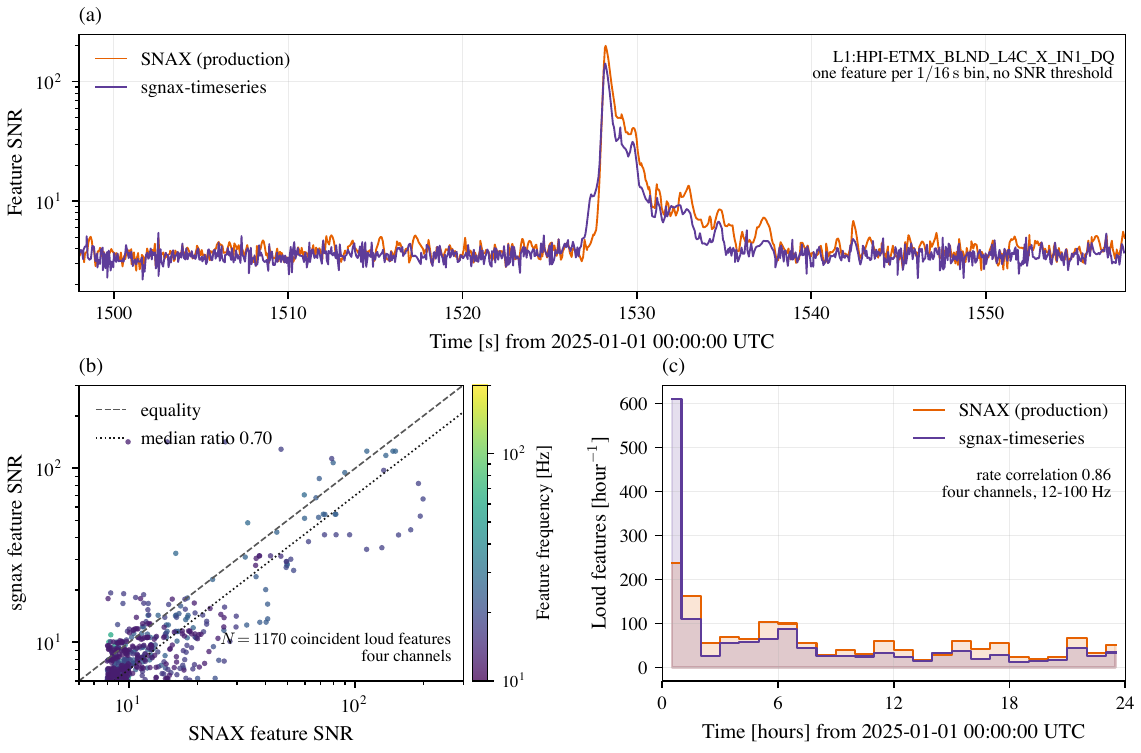}
  \caption{%
    The \texttt{sgnax-timeseries} feature mode against the \textsc{snax} production features, for four L1 auxiliary channels over the 24-hour interval beginning 2025-01-01 00:00:00 UTC.
    (a)~Sixty seconds of the dense $16\,\mathrm{Hz}$ feature stream on the HEPI L4C inertial-sensor channel: one feature per bin with no SNR threshold, tracing the noise floor and a loud scattered-light transient recovered by both pipelines.
    (b)~Recovered SNR of coincident loud features (production $\rho > 8$ in the common $12$--$100\,\mathrm{Hz}$ band, \texttt{sgnax} counterpart above its own $99$th percentile within one grid bin), colored by feature frequency; the dotted line marks the median ratio $0.70$, whose band structure and origin---a $\sqrt{2}$ over-amplification of the production rate bands below $25.6\,\mathrm{Hz}$---are established in the text.
    (c)~Hourly loud-feature rate summed over the four channels, under per-channel count-matched selections; the series correlate at $0.86$.%
  }
  \label{fig:timeseries_comparison}
\end{figure*}

Figure~\ref{fig:timeseries_comparison} summarizes the comparison.
Panel~(a) demonstrates the mode itself: both pipelines emit one feature per $62.5\,\mathrm{ms}$ bin with no SNR threshold, so the streams trace the full noise floor and rise together through a loud scattered-light transient, which both localize to the same time and frequency.
The populations agree well.
Of the $1470$ production features with $\rho > 8$ in the common band, $80\%$ ($73$--$92\%$ per channel) have a coincident \texttt{sgnax-timeseries} counterpart above that stream's own $99$th percentile within one grid bin; the frequencies of the coincident pairs correlate at Spearman rank $0.60$--$0.91$ per channel; and the hourly loud-feature rates of the two pipelines, compared under per-channel count-matched selections, correlate at $0.86$ across the day (panel c).
The per-feature time offsets of the coincident pairs have a median of $-8\,\mathrm{ms}$, within one grid bin; the tails are broader than the strain-channel comparison because the underlying scattered-light transients persist for seconds and the two banks, whose template floors differ, need not place the loudest template at the same instant of an extended event.

The one substantive residual is an SNR-scale difference, visible in panel~(b): on coincident loud features \texttt{sgnax-timeseries} recovers a median $0.70$ of the production SNR, while the two noise floors agree to $5$--$10\%$---an offset that persists with the template grids matched and therefore reflects a small difference in whitening response between the two chains rather than a bank effect.
This single scale difference drives all four \textsc{snax} residuals of Table~\ref{tab:residuals}: a fixed $\rho > 8$ selection on both streams yields a trigger-rate ratio of $0.46$, and the SNR-distribution KS statistic of $0.49$ reflects the shifted loud tail.
To determine which pipeline's scale is correct we calibrated the \texttt{sgnax} chain directly: sine-Gaussian injections of analytically known optimal SNR---into white noise and into colored noise with a steep $f^{-8}$ low-frequency wall mimicking the seismic spectra of these channels, at central frequencies down to $12.7\,\mathrm{Hz}$ and through the full whiten-resample-correlate path---are recovered to within a few percent of their injected values, with per-template noise response of unit variance.
The \texttt{sgnax} scale is thus the textbook matched-filter normalization of Eqs.~\eqref{eq:matched_filter}--\eqref{eq:snr_iq}, and the comparison implies that the production \textsc{snax} deployment reports SNRs $\sim\!1.4\times$ that normalization at low frequency.
A static audit of the production chain---template normalization, quadrature SNR assembly, multi-rate resampling corrections, and whitener normalization, each of which is individually correct in isolation---did not localize the excess; its origin is instead revealed by the band structure of the residual.
The ratio is not a smooth function of frequency but steps at the octave rate-band boundaries of the production pipeline.
For coincident features in the $12.8$--$25.6\,\mathrm{Hz}$ band---the nominal $64\,\mathrm{Hz}$ rate band, hosting $68\%$ of the loud pairs---the median ratio is $0.705$, equal to $1/\sqrt{2}$ to $0.3\%$, with per-sub-band medians within $4\%$ of $1/\sqrt{2}$ across the band.
Above $25.6\,\mathrm{Hz}$ the coincident pairs are sparse but show no comparable excess, and the production noise floor, which is flat in frequency for \texttt{sgnax}, steps at exactly $25.6\,\mathrm{Hz}$ in all four channels.
This is the signature of a multiband amplitude error in the production deployment: \textsc{snax} splits each whitened channel into octave rate bands and compensates the variance lost in downsampling with an amplitude gain computed for the band's nominal rate, but bands below the pipeline's $128\,\mathrm{Hz}$ resampling floor are in fact resampled only to that floor.
The nominal-$64\,\mathrm{Hz}$ band is therefore over-amplified by $\sqrt{128/64} = \sqrt{2}$, and the nominal-$32\,\mathrm{Hz}$ band, holding the templates below $12.8\,\mathrm{Hz}$, by a factor of $2$---the latter partially offset by the production $12\,\mathrm{Hz}$ high-pass filter, whose transition band overlaps that band and broadens its observed ratio distribution.
Because the excess is a pure amplitude scale on the affected bands---the measured ratio is independent of the distance of each production feature from the nearest \texttt{sgnax} template---it rescales SNRs without reordering them within a band, so threshold-based and rank-based consumers of the production features are largely unaffected in their selections, consistent with the population-level agreement above; absolute production SNRs below $25.6\,\mathrm{Hz}$, however, should be corrected by $1/\sqrt{2}$ before being compared or combined with \texttt{sgnax} features.

\subsection{A streaming glitch-rate frame channel}\label{sec:results:rate_channel}

Detector-characterization consumers frequently need the glitch \emph{rate} of a channel as a regularly sampled time series: it is the quantity tracked by detector summary pages, correlated against environmental and instrumental monitors, and consumed by statistical inference frameworks such as \textsc{idq}~\cite{essick2021idq}.
With the predecessor pipelines this product is derived offline, by histogramming trigger files after the fact, as was done for Fig.~\ref{fig:glitch_rate}.
\textsc{sgnax} instead emits it directly from the running pipeline as a native data product.
When either entry point is invoked with \texttt{-{}-save-glitch-rate}, a \texttt{TriggerRate} transform taps the channel's final trigger stream---the time-clustered stream of \texttt{sgnax-qtransform} or the window-aggregated stream of \texttt{sgnax-extract}---and bins it into a regularly sampled rate series, which a \textsc{gwf} sink element from \textsc{sgn-gwframe}~\cite{sgngwframe_repo} writes to standard frame files as a channel named \texttt{\{IFO\}:\{CHANNEL\}\_GLITCH\_RATE}.
Each sample records the instantaneous trigger rate in one bin, $62.5\,\mathrm{ms}$ wide at the default $16\,\mathrm{Hz}$ sample rate, and the series is contiguous and gap-free by construction: bins containing no triggers record a rate of zero, and offline runs cover exactly the requested GPS interval.
The selection entering the rate is itself configurable: \texttt{-{}-glitch-rate-snr-threshold} and \texttt{-{}-glitch-rate-frequency-range} restrict the counted triggers in SNR and central frequency without affecting the trigger files, so the channel can track a band of operational interest while the full trigger record is preserved.
Because the output is an ordinary frame channel, it can be read with any IGWN frame tooling and ingested alongside the detector channels themselves, with no trigger-file parsing required of the consumer.

\begin{figure*}[tp]
  \centering
  \includegraphics[width=0.98\textwidth]{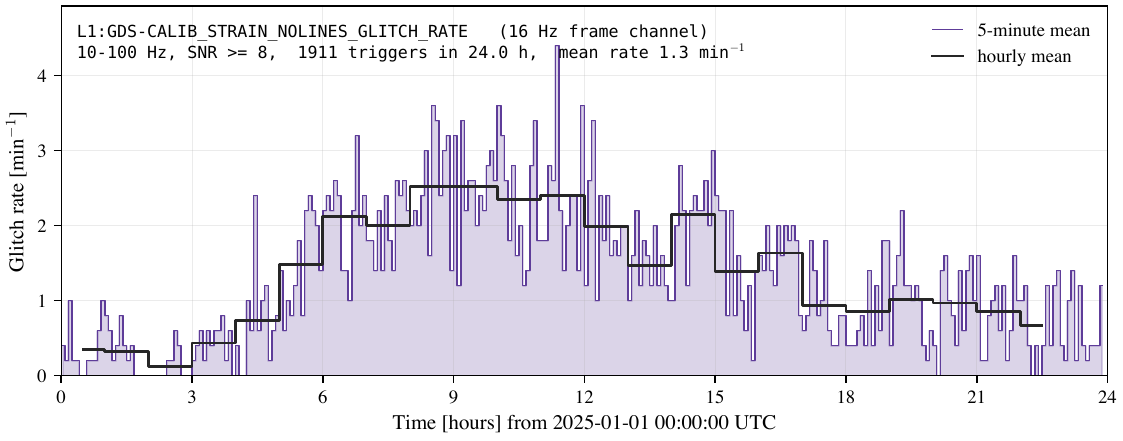}
  \caption{%
    The streaming glitch-rate frame channel written by \texttt{sgnax-qtransform -{}-save-glitch-rate} for the 24-hour \texttt{L1:GDS\_CALIB\_STRAIN\_NOLINES} interval of Sec.~\ref{sec:results:omicron}, with the rate restricted to the $10$--$100\,\mathrm{Hz}$, $\mathrm{SNR} \geq 8$ selection of Fig.~\ref{fig:glitch_rate}, shown in $5$-minute bins with the hourly mean overlaid (per-minute units, for direct comparison with Fig.~\ref{fig:glitch_rate}).
    Unlike the rate series of Fig.~\ref{fig:glitch_rate}, which is histogrammed offline from the trigger files, this series is the pipeline's own output, plotted exactly as read back from the \textsc{gwf} frames it wrote; it records $1911$ triggers and traces the same sustained daytime rate elevation.%
  }
  \label{fig:glitch_rate_channel}
\end{figure*}

Figure~\ref{fig:glitch_rate_channel} demonstrates the product on the same channel and 24-hour interval as the \textsc{omicron} comparison of Sec.~\ref{sec:results:omicron}, re-analyzed by \texttt{sgnax-qtransform} with the rate channel restricted to the $10$--$100\,\mathrm{Hz}$, $\mathrm{SNR} \geq 8$ selection of Fig.~\ref{fig:glitch_rate}.
The two figures show the same quantity but differ in provenance: the rate series of Fig.~\ref{fig:glitch_rate} was assembled by hand, by selecting and histogramming the trigger files offline after the analysis completed, whereas Fig.~\ref{fig:glitch_rate_channel} is the pipeline's own output, plotted exactly as read back from the frames it wrote.
The channel records $1911$ triggers and traces the same sustained daytime rate elevation, rising from a few tenths of a trigger per minute in the first hours of the run to a sustained $2$--$2.5\,\mathrm{min^{-1}}$ between roughly hours $6$ and $17$.
What the frame channel adds over the hand-derived rate is availability as a continuously sampled detector channel, with no offline post-processing required of the consumer and at low latency in the \texttt{devshm} and \textsc{arrakis} modes, suitable for summary-page trends, lock-loss and environmental correlation studies, and as a feature input to inference frameworks on the same cadence as the frames themselves.

\subsection{Runtime}\label{sec:results:runtime}

Wall-clock runtimes for the validation run of Table~\ref{tab:run_config} are reported in Table~\ref{tab:runtime}, separately for the matched-filter and Q-transform pipelines.
Both modes were benchmarked on the 24-hour \texttt{L1:GDS\_CALIB\_STRAIN\_NOLINES} interval of Sec.~\ref{sec:results:omicron}---$86{,}400$ channel-seconds of $16\,\mathrm{kHz}$ strain read from gravitational-wave frames---so the figures are directly comparable across modes.

\texttt{sgnax-qtransform} processes the full day in $14.1\,\mathrm{min}$ of wall-clock time: $9.8\,\mathrm{ms}$ per channel-second, or $102\times$ faster than real time.
Its CPU time ($13.2\,\mathrm{min}$) closely tracks its wall-clock time, so the Q-transform is essentially single-threaded in this configuration and leaves the remaining cores of a node free for additional channels.
\texttt{sgnax-extract} processes the same day in $84.0\,\mathrm{min}$ of wall clock---$58\,\mathrm{ms}$ per channel-second, or $17\times$ real time---distributing $8.1\,\mathrm{h}$ of CPU time across roughly six cores.
The higher cost of the matched-filter mode reflects its multi-rate template bank, which correlates several hundred templates at each of five sample rates (Sec.~\ref{sec:algorithms:matched_filter}) against every input sample, whereas the Q-transform evaluates a comparatively small set of $Q$-planes once per analysis chunk.

A direct \texttt{time} of the \textsc{omicron} executable on the same 24-hour strain interval---the apples-to-apples counterpart of the \textsc{sgnax} measurements above---completes in $16.0\,\mathrm{min}$ of wall clock, or $11\,\mathrm{ms}$ per channel-second and $90\times$ real time.
Its CPU time ($15.7\,\mathrm{min}$) likewise tracks its wall clock, so \textsc{omicron} too is essentially single-threaded on this input.
On this direct comparison the two excess-power pipelines agree to within $15\%$---\texttt{sgnax-qtransform} is in fact modestly faster than \textsc{omicron} ($14.1$ against $16.0\,\mathrm{min}$)---while the matched-filter \texttt{sgnax-extract} runs about $5\times$ slower than the \textsc{omicron} core, the expected cost of correlating a multi-rate template bank against every sample rather than projecting onto a handful of $Q$-planes once per analysis chunk.
This headroom over real time---$17\times$ for the matched filter and $102\times$ for the Q-transform---is what makes the low-latency operation discussed in Sec.~\ref{sec:discussion} practical: a pipeline that outruns the data by more than an order of magnitude has ample budget to keep pace with frames as they are written.
No batch wall-clock figure is quoted for \textsc{snax}: its production deployment operates as a persistent streaming service that consumes data at the acquisition rate by design, so the operative performance figure is its end-to-end latency, which Sec.~\ref{sec:results:latency} shows \texttt{sgnax-extract} matches.

\begin{table}[t]
  \caption{%
    Wall-clock runtime to analyze one 24-hour day ($86{,}400$ channel-seconds) of a single channel, measured on \texttt{L1:GDS\_CALIB\_STRAIN\_NOLINES} for 2025-01-01.
    The \textsc{sgnax} and \textsc{omicron} rows are direct \texttt{time} measurements on this interval.
    \textsc{snax} is deployed as a persistent streaming service rather than a batch job, so its operative performance figure is the end-to-end latency compared in Sec.~\ref{sec:results:latency}.%
  }
  \label{tab:runtime}
  \begin{tabular}{lcc}
    \hline\hline
    Pipeline & Wall clock & vs.\ real time \\
    \hline
    \texttt{sgnax-extract}                 & $84.0$\,min & $17\times$ \\
    \texttt{sgnax-qtransform}              & $14.1$\,min & $102\times$ \\
    \textsc{omicron}                       & $16.0$\,min & $90\times$ \\
    \textsc{snax}                          & \multicolumn{2}{c}{streaming (Sec.~\ref{sec:results:latency})} \\
    \hline\hline
  \end{tabular}
\end{table}

\subsection{Multi-channel scaling}\label{sec:results:scaling}

\begin{figure*}[tp]
  \centering
  \includegraphics[width=0.82\textwidth]{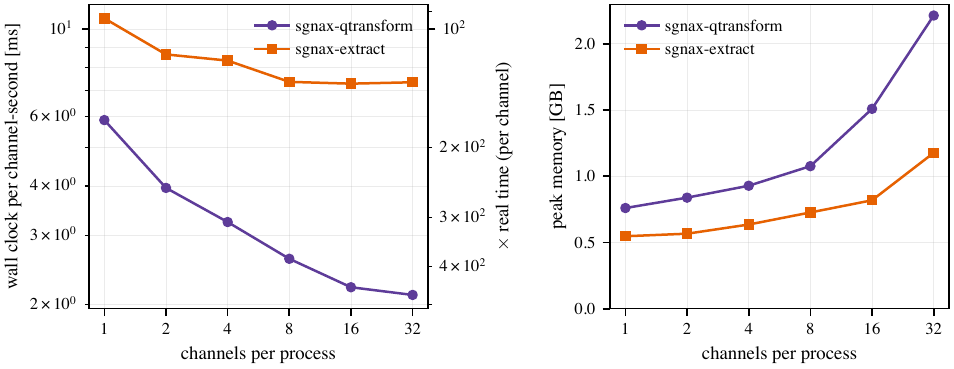}
  \caption{%
    Multi-channel throughput of the two \textsc{sgnax} pipelines, as a function of the number of channels analyzed as parallel branches of a single dataflow graph.
    Left: wall-clock cost per channel-second (left axis) and the equivalent per-channel real-time multiple (right axis), on log--log axes.
    Right: peak resident memory.
    Both are benchmarked over $1024\,\mathrm{s}$ of synthetic $4096\,\mathrm{Hz}$ channels on an Apple~M1~Pro, single process.
    The per-channel cost decreases as channels are added---the amortization of the shared source, whitening, and vectorized projection across channels---rather than remaining flat as it would for independent per-channel jobs.%
  }
  \label{fig:channel_scaling}
\end{figure*}

The single-channel figures of Table~\ref{tab:runtime} understate the throughput of the pipeline, because the defining architectural feature of \textsc{sgnax} is that an arbitrary list of channels is analyzed as parallel branches of one dataflow graph (Sec.~\ref{sec:architecture:topology}) rather than as independent per-channel jobs.
Figure~\ref{fig:channel_scaling} measures the effect: the wall-clock cost \emph{per channel-second} as a function of the number of channels carried through a single process, benchmarked over $1024\,\mathrm{s}$ of synthetic $4096\,\mathrm{Hz}$ auxiliary channels on an Apple~M1~Pro (the single-channel costs here are accordingly lower than those of Table~\ref{tab:runtime}, which analyzes the $16\,\mathrm{kHz}$ strain channel).
For \texttt{sgnax-qtransform} the per-channel cost falls from $5.9\,\mathrm{ms}$ per channel-second at one channel to $2.1\,\mathrm{ms}$ at $32$ channels---a $2.8\times$ amortization, or roughly $470\times$ real time per channel---while peak resident memory grows to only $2.2\,\mathrm{GB}$.
For \texttt{sgnax-extract} the amortization is more modest: the per-channel cost falls from $10.7\,\mathrm{ms}$ at one channel to $7.3\,\mathrm{ms}$ by eight and holds near there through $32$ channels---a $1.5\times$ reduction, limited by the heavier per-channel correlation load of the multi-rate template bank---while peak memory grows to $1.2\,\mathrm{GB}$ at $32$ channels.
A per-channel cost that \emph{decreases} as channels are added is the signature of shared work---a single source read, a single whitening front end, and per-$Q$-plane projections that vectorize across channels---and is precisely what the single-graph design buys over independent per-channel jobs, whose per-channel cost would instead be flat.
This headroom is what allows a single process to keep pace with a substantial channel count at low latency, and it sets the budget for the online operation discussed in Sec.~\ref{sec:discussion}.

In production the full auxiliary-channel set---of order $10^{3}$ channels per detector~\cite{robinet2020omicron}---is covered by distributing the channel list across \textsc{htcondor} jobs (Sec.~\ref{sec:algorithms:matched_filter}), each analyzing several tens of channels through one graph.
A deployment that analyzes, say, $32$ channels per process therefore covers the full set in of order $30$ jobs, and the per-process amortization of Fig.~\ref{fig:channel_scaling} means that each job clears its channels with a smaller core count than the same channels would require as independent per-channel jobs; the channel-to-job assignment is a free parameter, traded off against the per-job memory of Fig.~\ref{fig:channel_scaling} and the granularity desired for job-level failure recovery.
Section~\ref{sec:results:gpu} measures throughput above real time for nearly the full production channel list using four processes sharing one accelerator over a finite archival benchmark.

\subsection{GPU throughput}\label{sec:results:gpu}

We supplement the CPU scaling study with a single-device benchmark of \texttt{sgnax-qtransform} on one NVIDIA Tesla V100 ($16\,\mathrm{GB}$) in a DGX-1 node (dual $20$-core Xeon E5-2698~v4, $512\,\mathrm{GB}$).
The workload is $512\,\mathrm{s}$ of archival L1 data beginning at GPS $1419724818$, with auxiliary channels drawn from the $794$-channel O4 \textsc{omicron} production list at their native rates ($256\,\mathrm{Hz}$--$16.384\,\mathrm{kHz}$), $Q \in [\sqrt{11},150]$, and maximum mismatch $0.35$.
Channels are pre-extracted into compact frames to limit archival frame-reading overhead; the streaming real-time factor excludes one-time setup, and each successful configuration is summarized by the median of two runs.

\begin{figure*}[tp]
  \centering
  \includegraphics[width=0.82\textwidth]{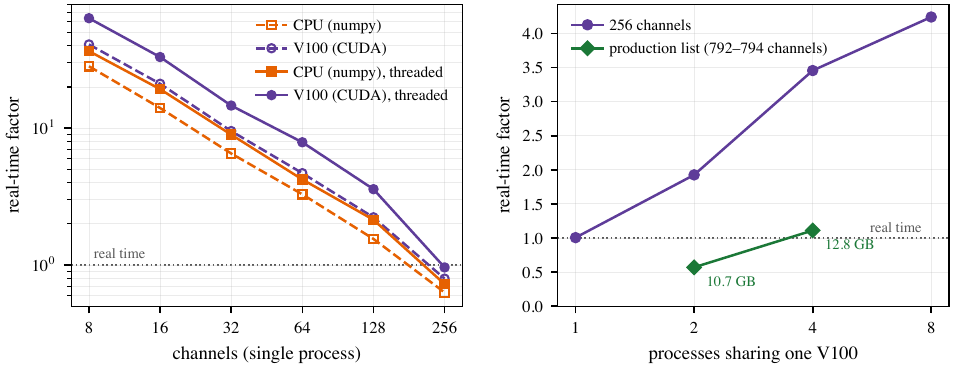}
  \caption{%
    Capacity of a single NVIDIA V100 for \texttt{sgnax-qtransform}, benchmarked on $512\,\mathrm{s}$ of archival L1 raw frames with channels drawn from the O4 \textsc{omicron} production list at their native sampling rates (Sec.~\ref{sec:results:gpu}).
    Left: streaming real-time factor versus the number of channels analyzed by a single process, for the numpy (CPU) and CUDA execution paths, without (open markers, dashed) and with (filled markers, solid) element-level threading; a single process saturates just below real time at $256$ channels with the GPU under $5\%$ utilized.
    Right: aggregate real-time factor when the channel list is partitioned into cost-balanced shards analyzed by independent processes sharing the one V100; $256$ channels reach $4.2\times$ real time at eight processes, and the four-process runs analyze $792$ channels at $1.11\times$ streaming real time using about $12.8$ of the $16\,\mathrm{GB}$ of device memory (annotations).%
  }
  \label{fig:gpu_capacity}
\end{figure*}

Figure~\ref{fig:gpu_capacity} shows the single-process and process-parallel measurements, all pinned to one device.
A single CUDA process reaches $0.96\times$ real time at $256$ channels with under $5\%$ mean GPU utilization; partitioning those channels among eight cost-balanced processes raises throughput to $4.2\times$ real time, consistent with host orchestration limiting single-process throughput.
The successful four-process runs analyze $792$ channels, with every shard above real time, at a median streaming factor of $1.11\times$, using about $12.8\,\mathrm{GB}$ of device memory and $5.6\%$ mean GPU utilization.
Including setup, these runs process $512\,\mathrm{s}$ in $485$ and $501\,\mathrm{s}$.
These measurements demonstrate throughput above real time for nearly the full production list on one GPU over the benchmark interval; continuous online capacity and production frame I/O are not measured here.

\subsection{Online latency}\label{sec:results:latency}

\begin{figure}[tp]
  \centering
  \includegraphics[width=0.92\columnwidth]{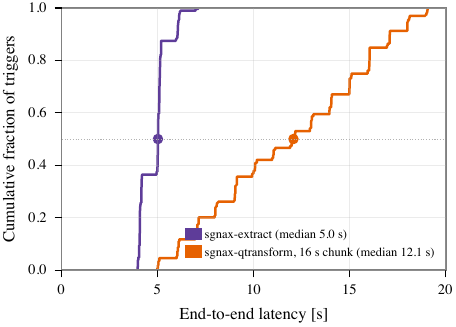}
  \caption{%
    Cumulative distribution of the measured end-to-end online latency---frame publication to trigger appearance---for the two \textsc{sgnax} modes reading a real-time one-second frame stream through the \texttt{devshm} backend of a mock shared-memory source.
    \texttt{sgnax-extract} sits at a chunk-free few-second floor set by the shared whitening front end; \texttt{sgnax-qtransform} (here a $16\,\mathrm{s}$ chunk) adds its chunk-fill time on top.
    Markers indicate the medians.
    Absolute values are indicative of a laptop mock; the chunk-free-versus-chunk-bound contrast is the robust result.%
  }
  \label{fig:latency}
\end{figure}

The latency argument of Sec.~\ref{sec:results:modes} rests on a structural claim---that the matched filter carries no per-chunk buffering while the Q-transform must fill a chunk---which we now measure directly.
We drive each pipeline through its \texttt{devshm} backend from a mock shared-memory source that publishes one-second frames in real time, and record, for each emitted trigger, the wall-clock interval between the publication of its frame and the appearance of the trigger in the output.
Figure~\ref{fig:latency} shows the resulting distributions.
\texttt{sgnax-extract} delivers triggers at a median end-to-end latency of $5.0\,\mathrm{s}$ ($6.3\,\mathrm{s}$ at the $99$th percentile), with no dependence on a chunk boundary; this floor is set by the shared whitening and acquisition front end rather than by the matched filter itself, and is comparable to the $\lesssim 5.5\,\mathrm{s}$ that the original \textsc{snax} achieved end-to-end~\cite{godwin2020thesis}, confirming that the reimplementation preserves the low-latency feature-generation capability that motivated \textsc{snax}.
\texttt{sgnax-qtransform}, run here with a $16\,\mathrm{s}$ chunk, instead shows a latency that ramps roughly uniformly across the chunk---median $12\,\mathrm{s}$, $99$th percentile $19\,\mathrm{s}$---as a trigger waits for the chunk enclosing it to fill; at the default $64\,\mathrm{s}$ chunk this floor rises to several tens of seconds, and it is tunable downward by shortening the chunk at the cost of coarser low-frequency resolution.
These absolute figures are indicative---they are measured against a mock \texttt{devshm} on a laptop, not a production deployment---but the structural contrast between the chunk-free matched filter and the chunk-bound Q-transform is the robust result, and it is what makes \texttt{sgnax-extract} the appropriate mode for feature generation on the cadence of the frames.

\section{Discussion}\label{sec:discussion}
\textsc{sgnax} is intended to occupy the same operational role currently filled by \textsc{snax} and \textsc{omicron} for transient extraction from strain and auxiliary channels in gravitational-wave detector characterization workflows~\cite{robinet2020omicron, mciver2019diagnostic, davis2021detchar}.
Both analysis modes accept strain or auxiliary channels, but their planned production roles differ.
We plan to run \texttt{sgnax-qtransform} on strain and the approximately $800$ auxiliary channels monitored by \textsc{omicron}, providing equivalent Q-transform trigger products through the shared pipeline.
The matched-filter mode has not been used operationally on strain; the archival strain runs reported here are validation and algorithm comparisons.
Its intended production role remains auxiliary-channel trigger and feature generation on shorter timescales, particularly for \textsc{idq}~\cite{essick2021idq, Huxford:2024vdt} and other consumers that require results faster than the chunk-based Q-transform can provide.
Beyond reproducing this role, the in-process \textsc{python} implementation also closes a long-standing latency gap: by consuming the same \texttt{/dev/shm} and \textsc{arrakis} sources used by low-latency analyses through the configuration object that drives offline runs, \texttt{sgnax-qtransform} can deliver Q-transform triggers to detector commissioners within tens of seconds of acquisition.
\textsc{omicron} itself supports an online mode with minutes-scale trigger delivery~\cite{robinet2020omicron}; in practice, however, the production deployments of recent observing runs have made triggers available to commissioners on the substantially longer cadences of the batch workflow, and it is this operational gap that the shared offline/online configuration of \textsc{sgnax} is designed to close.
Replacing two pipelines with a single \textsc{sgn}-based package reduces the operational surface area: a single configuration object, a single output schema, a single dependency tree, and a single set of monitoring hooks now suffice for analyses that previously spanned a \textsc{python}/\textsc{gstreamer} pipeline and a C++/\textsc{root} pipeline.
Adopting \textsc{sgnax} does require its triggers to be delivered to the downstream detector-characterization tools that today consume \textsc{snax} and \textsc{omicron} output, whether over the \textsc{kafka} streaming layer of Sec.~\ref{sec:architecture:workflow}, by a format adapter, or, more cleanly, by those tools standardizing on the common \textsc{hdf5} schema; the remaining per-consumer integration is beyond our scope here.
What matters for those consumers is that, irrespective of how the triggers are delivered, the data-quality information they receive is comparable to what the predecessors provided.
The injection-recovery and false-alarm studies of Sec.~\ref{sec:results} validate the \textsc{snax} matched-filter statistic that \texttt{sgnax-extract} inherits; the production comparison of Sec.~\ref{sec:results:omicron} reproduces the \textsc{omicron} trigger population to $0.1\%$ in total count; and the auxiliary-channel comparison of Sec.~\ref{sec:results:snax} shows the dense feature streams of \texttt{sgnax-timeseries} recovering the \textsc{snax} production loud-feature population with $80\%$ per-bin coincidence and hourly rates correlated at $0.86$.

The one substantive residual of that comparison is itself a product of the reimplementation.
Reproducing a pipeline against an independently calibrated reference is a more stringent test than reproducing it against its own output.
The injection calibration of Sec.~\ref{sec:results:snax} shows that the \texttt{sgnax} SNR scale is the textbook matched-filter normalization, and the band structure of the residual localizes the excess to a multiband amplitude error in the production \textsc{snax} deployment, which over-amplifies the rate bands below its $128\,\mathrm{Hz}$ resampling floor by $\sqrt{2}$ ($12.8$--$25.6\,\mathrm{Hz}$) and $2$ (below $12.8\,\mathrm{Hz}$).
Correcting the affected band by $1/\sqrt{2}$ brings the two pipelines' loud-feature SNR scales into agreement at the few-percent level.
Because the excess rescales SNRs without reordering them within a band, threshold- and rank-based consumers of the production features are unaffected in their selections; nonetheless, absolute production feature SNRs below $25.6\,\mathrm{Hz}$ carry the excess and should be corrected---or the upstream compensation fixed---before being mixed with \texttt{sgnax} features in a single training set.

\subsection{Outlook}\label{sec:discussion:outlook}

The \textsc{sgn} framework~\cite{huang2026sgn} provides a common foundation on which several IGWN tools---low-latency searches, detector-characterization pipelines, and offline reanalyses---can share elements.
\textsc{sgnax} is the first detector-characterization pipeline to be implemented natively on this foundation, and we expect additional detector-characterization tooling to follow.

In unifying the two pipelines, \textsc{sgnax} also delivers several capabilities that the predecessor designs identified as desirable but left to future work.
The matched-filter and excess-power statistics that McIver's event-trigger-generator study compared across separate, independently-configured tools~\cite{mciver2015thesis} are here available from a single configuration object, output schema, and clustering implementation, removing the cross-tool inconsistencies that complicated that comparison.
The streaming glitch-rate frame channel of Sec.~\ref{sec:results:rate_channel} supplies the regularly-sampled data-quality product anticipated for \textsc{snax}~\cite{godwin2020thesis}, and the Q-transform mode provides the multi-resolution, frequency-resolved view that the single aggregated matched-filter timeseries of the original \textsc{snax} could not.
The separate non-overlapping broadband power estimator takes a step toward the accurate total-SNR recovery of broadband bursts that~\cite{mciver2015thesis} flagged as an open problem (Sec.~\ref{sec:results:wnb}), though, as that section shows, the problem is only partially solved and remains a target for future work.
Other extensions anticipated by the predecessor designs remain natural next steps within the same framework: a probabilistic ranking statistic---the likelihood that a feature reflects non-stationary noise, rather than raw SNR~\cite{godwin2020thesis}---and a heterogeneous, glitch-informed template bank, including chirping waveforms, to target specific transient-noise morphologies rather than the homogeneous sine-Gaussian grid used here.

In summary, \textsc{sgnax} unifies the \textsc{snax} matched-filter and \textsc{omicron} excess-power statistics within a single \textsc{python}-native dataflow pipeline, with native multi-channel processing, three data-source backends spanning offline and low-latency operation, and a single DAG generator covering both tiled offline runs and long-lived \textsc{kafka}-mediated online deployments.
A process-parallel benchmark on one V100 measures throughput above real time for $792$ production auxiliary channels over a $512\,\mathrm{s}$ interval (Sec.~\ref{sec:results:gpu}).
On archival strain data, \texttt{sgnax-qtransform} reproduces the production \textsc{omicron} trigger population to $0.1\%$ in total count, with per-trigger SNRs matched at the percent level in stationary noise, including across the $10$--$100\,\mathrm{Hz}$ band that hosts scattered light and other transient-noise classes of central importance to detector characterization~\cite{davis2021detchar}; on production auxiliary channels \texttt{sgnax-timeseries} reproduces the \textsc{snax} feature population at matched cadence with an injection-calibrated SNR scale.
The residual differences are localized and mechanistically understood on both fronts: against \textsc{omicron}, a documented difference in whitening adaptation during sustained non-stationarity (Sec.~\ref{sec:results:omicron}); against \textsc{snax}, a multiband amplitude error in the production deployment, for which we report the mechanism and correction (Sec.~\ref{sec:results:snax}).
The accompanying open-source release~\cite{sgnax_repo} provides the implementation, validation suite, and documentation needed to migrate existing detector-characterization workflows.

\begin{acknowledgments}
The authors thank the developers of \textsc{snax}, \textsc{omicron}, and the \textsc{sgn} framework for the underlying tools on which this work builds, and the LIGO Scientific Collaboration detector-characterization working group for ongoing input on requirements and validation.
Z.Y. and G.G. acknowledge support from NSF Grant No.~PHY-2409740.
This work was supported by a grant from the Simons Foundation International [SFI-MPS-SSRFA-00023625, DD].
This material is based upon work supported by NSF's LIGO Laboratory which is a major facility fully funded by the National Science Foundation.
LIGO was constructed by the California Institute of Technology and Massachusetts Institute of Technology with funding from the National Science Foundation and operates under Cooperative Agreement PHY-2309200.
The authors are grateful for computational resources provided by the LIGO Laboratory and supported by National Science Foundation Grants PHY-0757058 and PHY-0823459.
LIGO Data Grid computing for LIGO detector characterization was supported by National Science Foundation awards 2110594 and 2513358.
This research has made use of data, software and/or web tools obtained from the LIGO Scientific Collaboration.
\end{acknowledgments}

\appendix
\section{Software implementation}\label{sec:appendix:implementation}

This appendix collects the software-engineering details of \textsc{sgnax} and of the predecessor pipelines that are referenced in the main text but are not essential to the algorithmic content.

\subsection{The \textsc{snax} and \textsc{omicron} implementations}\label{sec:appendix:implementation:predecessors}

The \textsc{snax} pipeline is implemented as a \textsc{python} layer over \textsc{gstreamer}-1.0~\cite{snax_repo, cannon2021gstlal}.
Each invocation of the \texttt{snax\_extract} command builds a \textsc{gstreamer} graph in which a \texttt{lalcachesrc} or live-data source feeds a \texttt{framecppchanneldemux}, then per-channel \texttt{whiten}, \texttt{audioresample}, and \texttt{lal\_matrixmixer} elements distribute samples across rate-specific filter banks before terminating in \texttt{appsink} callbacks; the \texttt{MultiChannelHandler} class collects sliced trigger rows into a \textsc{hdf5} or \textsc{kafka} sink.
Multi-channel coverage is achieved at the workflow level: \texttt{snax\_workflow} partitions the channel list among parallel \textsc{htcondor} jobs (each running a separate pipeline) and a downstream \texttt{snax\_combine} job concatenates the output \textsc{hdf5} files.
The dependence on \textsc{gstlal} couples \textsc{snax} to the entire \textsc{gstreamer} ABI~\cite{cannon2021gstlal}, so that upstream \textsc{gstreamer} releases can require compensating changes in the pipeline layer (e.g., a workaround for a memoryview regression in \texttt{gst-python}~1.18).

\textsc{omicron} depends on \textsc{root}~\cite{root1997} for I/O, plotting, and class serialization; on a version-locked sibling library \textsc{gwollum}~\cite{gwollum_repo} for trigger management, segments, and PSD estimation; on \textsc{frameL} for frame I/O; and on \textsc{fftw3}~\cite{frigo2005fftw} for transforms.
The build system is a hybrid of legacy \textsc{cmt} configuration files and a modern \textsc{cmake} layer~\cite{omicron_repo}.
A wrapper layer, \textsc{pyomicron}~\cite{pyomicron_repo}, provides condor-friendly \textsc{python} ergonomics around the underlying executable but does not address the underlying dependency footprint.
Taken together, this dependency footprint and the dual build system are the principal reason the pipeline is difficult to deploy in new computing environments.

\subsection{Package, dependencies, and layout}\label{sec:appendix:implementation:package}

\textsc{sgnax} is provided as the \textsc{python} package \texttt{sgnax}, available at~\cite{sgnax_repo}.
The package requires \textsc{python}~3.11 or later and depends on the four \textsc{sgn} ecosystem packages~\cite{huang2026sgn, sgnligo_repo, sgnarrakis_repo}, on \texttt{numpy}, \texttt{scipy}, \texttt{torch}~\cite{paszke2019pytorch}, \texttt{h5py}, \texttt{gwpy}~\cite{gwpy}, \texttt{lalsuite}~\cite{lalsuite}, \texttt{igwn-ligolw}~\cite{igwn_ligolw}, \texttt{igwn-segments}, \texttt{ligo-scald}~\cite{ligoscald_repo} (a toolkit for gravitational-wave data visualization and monitoring), the \textsc{arrakis} client library, \texttt{gpstime}, and \texttt{pyyaml}; DAG generation additionally requires \texttt{ezdag}, isolated in an optional \texttt{condor} extra so that the search entry points carry no \textsc{htcondor} dependency.
The package is dynamically versioned from the underlying \texttt{git} state via \texttt{hatch-vcs} and is built with the modern \textsc{python} packaging standard; a locked, multi-stage container image is built in continuous integration for deployment to \textsc{htcondor} execution sites.
The source tree is organized into element categories that mirror the \textsc{sgn} dataflow taxonomy: \texttt{src/sgnax/sources/} contains the unified \texttt{DataSourceInfo} object with its backend dispatch logic, the \textsc{toml} channel-configuration parser, and the \textsc{kafka} trigger source; \texttt{src/sgnax/transforms/} contains \texttt{TorchPeakFinder}, \texttt{WindowedTriggerAggregator}, \texttt{QScan}, \texttt{TriggerClusterer}, and the \texttt{FeatureSynchronizer}; \texttt{src/sgnax/waveforms/} contains the \texttt{SineGaussianBank} class; and \texttt{src/sgnax/sinks/} contains the \textsc{hdf5} feature sink and the \textsc{kafka} trigger sink.
The console entry points live in \texttt{src/sgnax/bin/}, and the workflow layer of Sec.~\ref{sec:architecture:workflow} is implemented by the \texttt{config}, \texttt{dags}, \texttt{cache}, and \texttt{kafka} modules, which provide the \textsc{yaml} configuration schema, the \textsc{htcondor} layer construction, the per-tile frame-cache selection, and the shared \textsc{kafka} topic-naming conventions, respectively.

\subsection{Static analysis and test suite}\label{sec:appendix:implementation:tests}

The package enforces a suite of static and dynamic checks intended to catch regressions in either algorithmic correctness or data-source configuration.
Static analysis is provided by \texttt{mypy} in strict mode, including \texttt{check\_untyped\_defs}, and by \texttt{ruff} with a broad rule set covering security, complexity, and import hygiene.
Dynamic testing is provided by \texttt{pytest}, with line coverage measured on every run.
The test suite includes twenty-five test modules covering the template-bank generator, the matched-filter peak finder and its timing conventions, the cross-rate aggregator, the Q-transform tiling and clustering, the data-source and channel-configuration objects, the \textsc{hdf5} feature sink and its \textsc{gwpy} compatibility, the \textsc{kafka} source, sink, and end-to-end round trip, the \textsc{yaml} configuration and DAG generator, the multi-rate whitening variance preservation, and an end-to-end correlation-SNR test that injects known-amplitude sine-Gaussians into white noise and verifies the recovered SNR.
A markdown-docs test plugin verifies that all code examples embedded in the documentation execute correctly, ensuring that documentation does not drift from the implementation.

\subsection{Workflow configuration and deployment}\label{sec:appendix:implementation:workflow}

Production analyses are deployed through a single DAG generator, \texttt{sgnax-dagger}, which builds \textsc{htcondor} workflows via \texttt{ezdag}; the dependency is isolated in an optional \texttt{condor} extra so that the search entry points themselves carry no \textsc{htcondor} requirement.
In offline mode the requested span is tiled into day-scale jobs aligned to a day grid and, when a segments file is supplied, clipped to observing segments; each job is pointed at a per-tile frame cache so that \textsc{htcondor} transfers only the frame files its tile overlaps, and per-channel-bin \texttt{sgnax-merge} jobs concatenate the tile outputs into one final feature file per bin.
In online mode the generator instead launches long-lived search jobs against a low-latency source, each publishing its trigger stream to a per-job \textsc{kafka} topic; a \texttt{sgnax-synchronize} job merges the per-job streams into a single time-ordered stream on a shared topic, and a \texttt{sgnax-archive} job consumes that stream and writes rotating feature files in the schema of Sec.~\ref{sec:architecture:triggers}.
Topic names are derived on both the publishing and consuming sides from a per-analysis identifier, so no topic is spelled out by hand.
Every option of the generator and of the underlying entry points can equivalently be supplied through a \textsc{yaml} configuration file, with command-line flags taking precedence; per-channel search parameters live in the \textsc{toml} channel configuration shared with the search entry points.
A locked, multi-stage container image built in continuous integration carries the pipeline and a source checkout to the execution sites, so that the generator plays the operational role of the \texttt{snax\_workflow}/\texttt{snax\_combine} tooling of \textsc{snax} and the \textsc{pyomicron} wrapper of \textsc{omicron}.

\section{Mismatch-driven bank spacing}\label{sec:appendix:bank_spacing}

This appendix recapitulates the mismatch-driven bank-spacing rules used by both the matched-filter pipeline of Sec.~\ref{sec:algorithms:matched_filter} and the Q-transform pipeline of Sec.~\ref{sec:algorithms:qtransform}.
The derivations follow the thesis of Chatterji~\cite{chatterji2005} and the technical note of Robinet~\cite{robinet2018technote}; we restate the rules here for self-containment and for direct comparison with the corresponding constants in the \textsc{sgnax} source.

The squared distance between two nearby points $(\tau, f, Q)$ and $(\tau + \delta\tau, f + \delta f, Q + \delta Q)$ in the parameter space, expressed in terms of the mismatch metric, is
\begin{equation}\label{eq:metric}
  \delta s^2
    = \frac{4\pi^2 f^2}{Q^2}\,\delta\tau^2
    + \frac{2 + Q^2}{4 f^2}\,\delta f^2
    + \frac{1}{2 Q^2}\,\delta Q^2.
\end{equation}
A grid that places templates at squared distance $\delta s^2 \leq \mu_{\max}/3$ along each independent direction guarantees that the worst-case mismatch between an arbitrary signal in the parameter space and the nearest template is at most $\mu_{\max}$.

Integrating Eq.~\eqref{eq:metric} along the $Q$ direction at fixed $f$ gives the number of $Q$ planes needed to span $[Q_{\min}, Q_{\max}]$,
\begin{equation}\label{eq:nq}
  N_Q = \left\lceil
    \frac{1}{2\sqrt{\mu_{\max}/3}}\,
    \frac{1}{\sqrt{2}}\,
    \ln\!\left(\frac{Q_{\max}}{Q_{\min}}\right)
  \right\rceil,
\end{equation}
in agreement with Eq.~\eqref{eq:num_q_planes}.
At fixed $Q$, the number of frequency bands required to span $[f_{\min}, f_{\max}]$ is
\begin{equation}\label{eq:nf}
  N_f(Q) = \left\lceil
    \frac{1}{2\sqrt{\mu_{\max}/3}}\,
    \frac{\sqrt{2 + Q^2}}{2}\,
    \ln\!\left(\frac{f_{\max}}{f_{\min}}\right)
  \right\rceil,
\end{equation}
so that high-$Q$ planes have a denser frequency grid than low-$Q$ planes, reflecting the narrower bandwidth $\delta f/f \propto 1/Q$ of high-$Q$ tiles.
The two expressions are implemented in the \texttt{sgnax.waveforms} module (matched-filter bank) and the \texttt{QScan} transform (Q-transform bank); the same constants govern the \textsc{omicron} tile counts of Robinet~\cite{robinet2018technote}.

\section{Data-source backends}\label{sec:appendix:data_sources}

This appendix documents the three backends accepted by the \texttt{DataSourceInfo} object of Sec.~\ref{sec:architecture:sources}.
Each backend is selected by the \texttt{-{}-data-source} command-line argument and each carries its own set of subordinate options.
After source selection, every backend feeds the same \texttt{Whiten} stage, which produces each channel at its configured working rate, capped at its native input rate.
Whitening is enabled by default for all three backends and can be disabled with \texttt{-{}-no-whiten} for debugging or intermediate-data inspection.
The backend selector \texttt{devshm} refers to data read from the filesystem path \texttt{/dev/shm}.

\subsection{\texttt{frames}}\label{sec:appendix:data_sources:frames}

The \texttt{frames} backend reads from a frame-cache file (\texttt{.lcf}) listing GWF frames~\cite{lalsuite}.
Optional segment files can be supplied to gate the analysis on observatory state.
This backend is intended for offline analyses on archival data and is the most direct analog of the \texttt{snax\_extract -{}-data-source frames} mode.

\subsection{\texttt{devshm}}\label{sec:appendix:data_sources:devshm}

The \texttt{devshm} backend reads from a shared-memory ring buffer at \texttt{/dev/shm}, populated at the IGWN sites by the low-latency frame distribution.
A configurable source-queue timeout (\texttt{-{}-source-queue-timeout}) specifies how long the source will block on a missing frame before declaring a discontinuity.
This backend is intended for low-latency online deployments and corresponds to the \texttt{snax\_extract -{}-data-source devshm} mode.

\subsection{\texttt{arrakis}}\label{sec:appendix:data_sources:arrakis}

The \texttt{arrakis} backend reads from the \textsc{arrakis} timeseries distribution service~\cite{arrakis} via the \textsc{sgn-arrakis} source element~\cite{sgnarrakis_repo}.
\textsc{arrakis} is a new low-latency data-delivery service, currently deployed at the LIGO sites and intended to serve IGWN-wide distribution in future observing runs.
This backend is intended for low-latency online deployments and replaces the role of the \texttt{framexmit} multicast source used by \textsc{snax} prior to the \textsc{arrakis} migration.

\bibliography{references}

@article{aasi2015aligo,
  author  = {Aasi, J. and others},
  title   = {{Advanced LIGO}},
  journal = {Class. Quant. Grav.},
  volume  = {32},
  pages   = {074001},
  year    = {2015},
  doi     = {10.1088/0264-9381/32/7/074001},
  eprint  = {1411.4547},
}

@misc{abac2025gwtc4,
  author        = {Abac, A. and others},
  title         = {{GWTC-4.0: Updating the Gravitational-Wave Transient Catalog with Observations from the First Part of the Fourth LIGO-Virgo-KAGRA Observing Run}},
  eprint        = {2508.18082},
  archivePrefix = {arXiv},
  primaryClass  = {gr-qc},
  year          = {2025},
}

@misc{abac2026gwtc5,
  author        = {Abac, A. and others},
  title         = {{GWTC-5.0: Observations from the Second Part of the Fourth LIGO-Virgo-KAGRA Observing Run and Updates to the Gravitational-Wave Transient Catalog}},
  eprint        = {2605.27225},
  archivePrefix = {arXiv},
  primaryClass  = {gr-qc},
  year          = {2026},
}

@article{abbott2016,
  author  = {Abbott, B. P. and others},
  title   = {Observation of Gravitational Waves from a Binary Black Hole Merger},
  journal = {Phys. Rev. Lett.},
  volume  = {116},
  pages   = {061102},
  year    = {2016},
  doi     = {10.1103/PhysRevLett.116.061102},
  eprint  = {1602.03837},
}

@article{abbott2016detchar,
  author  = {Abbott, B. P. and others},
  title   = {{Characterization of transient noise in Advanced LIGO relevant to gravitational wave signal GW150914}},
  journal = {Class. Quant. Grav.},
  volume  = {33},
  pages   = {134001},
  year    = {2016},
  doi     = {10.1088/0264-9381/33/13/134001},
  eprint  = {1602.03844},
}

@article{abbott2021gwtc3,
  author  = {Abbott, R. and others},
  title   = {{GWTC-3: Compact Binary Coalescences Observed by LIGO and Virgo during the Second Part of the Third Observing Run}},
  journal = {Phys. Rev. X},
  volume  = {13},
  pages   = {041039},
  year    = {2023},
  doi     = {10.1103/PhysRevX.13.041039},
  eprint  = {2111.03606},
}

@article{acernese2015advirgo,
  author  = {Acernese, F. and others},
  title   = {{Advanced Virgo: a second-generation interferometric gravitational wave detector}},
  journal = {Class. Quant. Grav.},
  volume  = {32},
  pages   = {024001},
  year    = {2015},
  doi     = {10.1088/0264-9381/32/2/024001},
  eprint  = {1408.3978},
}

@article{anderson2001excesspower,
  author  = {Anderson, Warren G. and Brady, Patrick R. and Creighton, Jolien D. E. and Flanagan, \'Eanna \'E.},
  title   = {{An excess power statistic for detection of burst sources of gravitational radiation}},
  journal = {Phys. Rev. D},
  volume  = {63},
  pages   = {042003},
  year    = {2001},
  doi     = {10.1103/PhysRevD.63.042003},
  eprint  = {gr-qc/0008066},
}

@article{brown1991qtransform,
  author  = {Brown, Judith C.},
  title   = {{Calculation of a constant Q spectral transform}},
  journal = {J. Acoust. Soc. Am.},
  volume  = {89},
  pages   = {425--434},
  year    = {1991},
  doi     = {10.1121/1.400476},
}

@article{chatterji2004multiresolution,
  author  = {Chatterji, S. and Blackburn, L. and Martin, G. and Katsavounidis, E.},
  title   = {{Multiresolution techniques for the detection of gravitational-wave bursts}},
  journal = {Class. Quant. Grav.},
  volume  = {21},
  pages   = {S1809--S1818},
  year    = {2004},
  doi     = {10.1088/0264-9381/21/20/024},
  eprint  = {gr-qc/0412119},
}

@phdthesis{chatterji2005,
  author  = {Chatterji, Shourov K.},
  title   = {{The search for gravitational wave bursts in data from the second LIGO science run}},
  school  = {Massachusetts Institute of Technology},
  year    = {2005},
}

@article{davis2021detchar,
  author  = {Davis, D. and others},
  title   = {{LIGO detector characterization in the second and third observing runs}},
  journal = {Class. Quant. Grav.},
  volume  = {38},
  pages   = {135014},
  year    = {2021},
  doi     = {10.1088/1361-6382/abfd85},
  eprint  = {2101.11673},
}

@misc{davis2026dqr,
  author        = {Davis, Derek and Yarbrough, Zach and Areeda, Joseph and Macas, Ronaldas and Arnaud, Nicolas and Helmling-Cornell, Adrian and Doliva, Paolina and Godwin, Olivia and Yuzurihara, Hirotaka and Mannix, Benjamin and others},
  title         = {{Rapid data quality investigations of gravitational-wave events with the Data Quality Report Builder toolkit}},
  eprint        = {2605.16183},
  archivePrefix = {arXiv},
  year          = {2026},
}

@article{frigo2005fftw,
  author  = {Frigo, Matteo and Johnson, Steven G.},
  title   = {{The Design and Implementation of FFTW3}},
  journal = {Proc. IEEE},
  volume  = {93},
  pages   = {216--231},
  year    = {2005},
  doi     = {10.1109/JPROC.2004.840301},
}

@phdthesis{godwin2020thesis,
  author  = {Godwin, Patrick},
  title   = {{Low-latency Statistical Data Quality in the Era of Multi-Messenger Astronomy}},
  school  = {The Pennsylvania State University},
  year    = {2020},
}

@misc{gvt_repo,
  author       = {Godwin, Patrick},
  title        = {{gvt: a Glitch Validation Toolkit}},
  year         = {2024},
  howpublished = {\url{https://git.ligo.org/patrick.godwin/gvt}},
}

@article{gwpy,
  author  = {Macleod, Duncan M. and Areeda, Joseph S. and Coughlin, Scott B. and Massinger, Thomas J. and Urban, Alexander L.},
  title   = {{GWpy: A Python package for gravitational-wave astrophysics}},
  journal = {SoftwareX},
  volume  = {13},
  pages   = {100657},
  year    = {2021},
  doi     = {10.1016/j.softx.2021.100657},
}

@misc{igwn_ligolw,
  author       = {{IGWN Software Working Group}},
  title        = {{igwn-ligolw: LIGO Lightweight XML I/O for Python}},
  year         = {2024},
  howpublished = {\url{https://git.ligo.org/computing/software/igwn-ligolw}},
}

@article{glanzer2023gravityspy,
  author  = {Glanzer, J. and others},
  title   = {{Data quality up to the third observing run of advanced LIGO: Gravity Spy glitch classifications}},
  journal = {Class. Quant. Grav.},
  volume  = {40},
  pages   = {065004},
  year    = {2023},
  doi     = {10.1088/1361-6382/acb633},
  eprint  = {2208.12849},
}

@misc{glanzer2026o4bcdetchar,
  author       = {Glanzer, J. and Helmling-Cornell, A. F. and others},
  title        = {{LIGO Detector Characterization in the Second and Third Parts of the Fourth Observing Run}},
  year         = {2026},
  howpublished = {LIGO Document P2600357, in preparation},
}

@article{kagra2019,
  author  = {{KAGRA Collaboration}},
  title   = {{KAGRA: 2.5 generation interferometric gravitational wave detector}},
  journal = {Nature Astron.},
  volume  = {3},
  pages   = {35--40},
  year    = {2019},
  doi     = {10.1038/s41550-018-0658-y},
  eprint  = {1811.08079},
}

@misc{lalsuite,
  author       = {{LIGO Scientific Collaboration} and {Virgo Collaboration} and {KAGRA Collaboration}},
  title        = {{LVK Algorithm Library -- LALSuite}},
  year         = {2018},
  doi          = {10.7935/GT1W-FZ16},
  howpublished = {Free software (GPL)},
}

@phdthesis{mciver2015thesis,
  author  = {McIver, Jessica},
  title   = {{The impact of terrestrial noise on the detectability and reconstruction of gravitational wave signals from core-collapse supernovae}},
  school  = {University of Massachusetts Amherst},
  year    = {2015},
}

@incollection{mciver2019diagnostic,
  author    = {McIver, J. and Massinger, T. J. and Robinet, F. and Smith, J. R. and Walker, M.},
  title     = {{Diagnostic methods for gravitational-wave detectors}},
  booktitle = {Advanced Interferometric Gravitational-wave Detectors},
  editor    = {Reitze, D. and Saulson, P. and Grote, H.},
  volume    = {1},
  pages     = {373--392},
  publisher = {World Scientific},
  year      = {2019},
  doi       = {10.1142/9789813146082_0014},
}

@article{essick2021idq,
  author  = {Essick, Reed and Godwin, Patrick and Hanna, Chad and Blackburn, Lindy and Katsavounidis, Erik},
  title   = {{iDQ: Statistical inference of non-Gaussian noise with auxiliary degrees of freedom in gravitational-wave detectors}},
  journal = {Mach. Learn.: Sci. Technol.},
  volume  = {2},
  pages   = {015004},
  year    = {2021},
  doi     = {10.1088/2632-2153/abab5f},
}

@misc{huang2026sgn,
  author        = {Huang, Yun-Jing and Godwin, Olivia and Hanna, Chad and Kennington, James and Rollins, Jameson and Melching, Max and Sovitzky, Nathanael E. and Viets, Aaron and Wade, Madeline and Yarbrough, Zach and Chu, Yu-Kuang and Phillips, William Wyatt and Sachdev, Surabhi and Udall, Rhiannon},
  title         = {{SGN: A python framework for stream-processing pipelines}},
  eprint        = {2607.03575},
  archivePrefix = {arXiv},
  primaryClass  = {astro-ph.IM},
  doi           = {10.48550/arXiv.2607.03575},
  year          = {2026},
}

@misc{Huxford:2024vdt,
  author        = {Huxford, Rachael and George, Richard and Trevor, Max and Yarbrough, Zach and Godwin, Patrick},
  title         = {{Performance of iDQ ahead of LIGO, Virgo, and KAGRA's fourth observing run}},
  eprint        = {2412.04638},
  archivePrefix = {arXiv},
  primaryClass  = {gr-qc},
  reportNumber  = {LIGO-P2400455},
  month         = {12},
  year          = {2024},
}

@article{messick2017,
  author  = {Messick, Cody and others},
  title   = {{Analysis Framework for the Prompt Discovery of Compact Binary Mergers in Gravitational-wave Data}},
  journal = {Phys. Rev. D},
  volume  = {95},
  pages   = {042001},
  year    = {2017},
  doi     = {10.1103/PhysRevD.95.042001},
  eprint  = {1604.04324},
}

@inproceedings{paszke2019pytorch,
  author    = {Paszke, Adam and others},
  title     = {{PyTorch: An Imperative Style, High-Performance Deep Learning Library}},
  booktitle = {Advances in Neural Information Processing Systems 32},
  pages     = {8024--8035},
  year      = {2019},
}

@article{robinet2020omicron,
  author  = {Robinet, F. and Arnaud, N. and Leroy, N. and Lundgren, A. and Macleod, D. and McIver, J.},
  title   = {{Omicron: A tool to characterize transient noise in gravitational-wave detectors}},
  journal = {SoftwareX},
  volume  = {12},
  pages   = {100620},
  year    = {2020},
  doi     = {10.1016/j.softx.2020.100620},
  eprint  = {2007.11374},
}

@misc{robinet2018technote,
  author       = {Robinet, F.},
  title        = {{Omicron: an algorithm to detect and characterize transient events in gravitational-wave detectors}},
  year         = {2018},
  howpublished = {Virgo Technical Document VIR-0545C-14},
  url          = {https://tds.virgo-gw.eu/ql/?c=10651},
}

@misc{ligoscald_repo,
  author       = {Godwin, Olivia and others},
  title        = {{ligo-scald: SCalable Analytics for LVK Data}},
  howpublished = {\url{https://git.ligo.org/gstlal-visualisation/ligo-scald}},
  note         = {Accessed October 2026},
}

@misc{omicron_repo,
  author       = {Robinet, F. and others},
  title        = {{Omicron: a Q-transform analysis tool for gravitational-wave detector data}},
  year         = {2025},
  howpublished = {\url{https://git.ligo.org/virgo/virgoapp/Omicron}},
}

@misc{gwollum_repo,
  author       = {Robinet, F.},
  title        = {{GWOLLUM: a toolkit for analysis of gravitational-wave detector data}},
  year         = {2025},
  howpublished = {\url{https://git.ligo.org/virgo/virgoapp/GWOLLUM}},
}

@misc{pyomicron_repo,
  author       = {Macleod, D. and Urban, A.},
  title        = {{pyomicron: utilities for running and post-processing Omicron analyses}},
  year         = {2019},
  doi          = {10.5281/zenodo.2636552},
  howpublished = {\url{https://github.com/gwpy/pyomicron}},
}

@article{root1997,
  author  = {Brun, R. and Rademakers, F.},
  title   = {{ROOT: an object-oriented data analysis framework}},
  journal = {Nucl. Instrum. Meth. A},
  volume  = {389},
  pages   = {81--86},
  year    = {1997},
  doi     = {10.1016/S0168-9002(97)00048-X},
}

@misc{sgnligo_repo,
  author       = {{SGN Development Team}},
  title        = {{sgn-ligo: LIGO-specific elements for the SGN framework}},
  year         = {2026},
  howpublished = {\url{https://git.ligo.org/greg/sgn-ligo}},
}

@misc{sgnarrakis_repo,
  author       = {{SGN Development Team}},
  title        = {{sgn-arrakis: Arrakis data-source elements for the SGN framework}},
  year         = {2026},
  howpublished = {\url{https://git.ligo.org/ngdd/sgn-arrakis}},
}

@misc{sgnax_repo,
  author       = {Godwin, Olivia and Yarbrough, Zachary},
  title        = {{sgnax: a multi-channel auxiliary burst-search pipeline}},
  year         = {2026},
  howpublished = {\url{https://git.ligo.org/detchar/sgn-dq/sgnax}},
}

@misc{sgngwframe_repo,
  author       = {{SGN Development Team}},
  title        = {{sgn-gwframe: gravitational-wave frame file I/O elements for the SGN framework}},
  year         = {2026},
  howpublished = {\url{https://git.ligo.org/greg/sgn-gwframe}},
}

@misc{snax_repo,
  author       = {Godwin, Patrick and others},
  title        = {{snax: Stream-based Noise Acquisition and eXtraction}},
  year         = {2022},
  howpublished = {\url{https://git.ligo.org/snax/snax}},
}

@article{soni2021scattering,
  author  = {Soni, S. and others},
  title   = {{Reducing scattered light in LIGO's third observing run}},
  journal = {Class. Quant. Grav.},
  volume  = {38},
  pages   = {025016},
  year    = {2021},
  doi     = {10.1088/1361-6382/abc906},
  eprint  = {2007.14876},
}

@article{soni2025o4adetchar,
  author  = {Soni, S. and others},
  title   = {{LIGO Detector Characterization in the first half of the fourth Observing run}},
  journal = {Class. Quant. Grav.},
  volume  = {42},
  pages   = {085016},
  year    = {2025},
  doi     = {10.1088/1361-6382/adc4b6},
  eprint  = {2409.02831},
}

@misc{arrakis,
  author       = {{Arrakis Developers}},
  title        = {{arrakis-python: Python client for the Arrakis low-latency timeseries data distribution service}},
  year         = {2024},
  howpublished = {\url{https://git.ligo.org/ngdd/arrakis-python}},
}

@article{cannon2021gstlal,
  author  = {Cannon, Kipp and others},
  title   = {{GstLAL: A software framework for gravitational wave discovery}},
  journal = {SoftwareX},
  volume  = {14},
  pages   = {100680},
  year    = {2021},
  doi     = {10.1016/j.softx.2021.100680},
  eprint  = {2010.05082},
}

@article{zevin2017gravityspy,
  author  = {Zevin, M. and others},
  title   = {{Gravity Spy: integrating Advanced LIGO detector characterization, machine learning, and citizen science}},
  journal = {Class. Quant. Grav.},
  volume  = {34},
  pages   = {064003},
  year    = {2017},
  doi     = {10.1088/1361-6382/aa5cea},
  eprint  = {1611.04596},
}

@article{smith2011hveto,
  author  = {Smith, J. R. and Abbott, T. and Hirose, E. and Leroy, N. and Macleod, D. and McIver, J. and Saulson, P. and Shawhan, P.},
  title   = {{A hierarchical method for vetoing noise transients in gravitational-wave detectors}},
  journal = {Class. Quant. Grav.},
  volume  = {28},
  pages   = {235005},
  year    = {2011},
  doi     = {10.1088/0264-9381/28/23/235005},
  eprint  = {1107.2948},
}

@article{biswas2013mla,
  author  = {Biswas, R. and others},
  title   = {{Application of machine learning algorithms to the study of noise artifacts in gravitational-wave data}},
  journal = {Phys. Rev. D},
  volume  = {88},
  pages   = {062003},
  year    = {2013},
  doi     = {10.1103/PhysRevD.88.062003},
  eprint  = {1303.6984},
}

\end{document}